\documentclass[11pt,a4paper]{article}
\usepackage[utf8]{inputenc}
\usepackage[T1]{fontenc}
\usepackage{lmodern}
\usepackage[margin=2.5cm]{geometry}
\usepackage{amsmath,amssymb,bm}
\usepackage{graphicx}
\usepackage{booktabs,multirow}
\usepackage{setspace}
\usepackage[font=small,labelfont=bf]{caption}
\usepackage{float}
\usepackage[hidelinks]{hyperref}
\usepackage{xcolor}

\graphicspath{{figs/}}
\hypersetup{
  pdftitle={Damage Assessment of Supporting Pillars in an Underground Cave Using Joint Inversion of Electrical Resistivity and P-wave Velocity, Burgundy (France)},
  pdfauthor={A. Carrier, P. Bottelin, L. Fabre, A. Mathy},
  pdfsubject={Preprint},
  pdfkeywords={Joint inversion, refraction seismic tomography, electrical resistivity tomography, underground cave, structural health monitoring, clustering}
}

\begin{document}

\begin{center}
\fbox{\parbox{0.92\textwidth}{\small\setstretch{1.1}
\textbf{Preprint.} This is the authors' submitted version of the article published in
\emph{Pure and Applied Geophysics}, 179 (2022), 45--67.
The final version is available at \url{https://doi.org/10.1007/s00024-021-02921-w}.
This version may differ from the published version.}}
\end{center}

\vspace{1em}
\begin{center}
{\Large\bfseries Damage Assessment of Supporting Pillars in an Underground Cave Using Joint Inversion of Electrical Resistivity and P-wave Velocity, Burgundy (France)\par}
\vspace{1.2em}
{\large A.~Carrier$^{1,*}$, P.~Bottelin$^{1}$, L.~Fabre$^{2}$, A.~Mathy$^{2}$\par}
\vspace{0.8em}
{\small\setstretch{1.1}
$^{1}$Association pour le Développement des Recherches sur les Glissements de Terrain (ADRGT), Grenoble, France\\
$^{2}$Société Alpine de Géotechnique (SAGE), Grenoble, France\\
$^{*}$Corresponding author: a.carrier@adrgt.org\par}
\end{center}

\vspace{1em}
\section*{Abstract}
This study proposes the combined use of refraction seismic tomography and electrical resistivity tomography to assess the structural health of supporting pillars in an underground cave (Burgundy, France). A three-phases petrophysical joint inversion (P-JIv) was performed with a finite-element model on an unstructured triangular mesh. P-JIv showed better resolution and results in better agreement between geophysical methods than independent inversions. Results revealed contrasted electrical resistivity and P-wave velocity zones related to fracturing and weathering of rock. Geophysical results showed a very good agreement with visual observations and drilling logs. Rock porosity and water content were derived from P-JIv results. An automatic clustering process was applied to derive a proxy for pillar integrity, which was used to prioritize the cave reinforcement works.

\medskip
\noindent\textbf{Keywords:} Joint inversion, refraction seismics tomography, electrical resistivity tomography, underground cave, structural health investigation and monitoring, clustering, geophysics.

\section{Introduction}

The number of underground infrastructures has increased rapidly over the last decades, especially in dense urban areas (Broere, 2016). These underground cavities may consist of abandoned or functional quarries, natural caves, touristic sites, urban systems, underground plants, etc. Structural health investigation (SHI) and monitoring (SHM) techniques represent a crucial decision help for municipalities and operators in order to ensure safety of both underground and surface facilities (Al Heib et al., 2015; Brownjohn, 2007). SHI may be based on geological surveys, destructive geotechnical campaigns (e.g. drillings logs, corings), microstructural investigations, geomechanical investigations and modeling (e.g. tributary area method, Brady and Brown, 2006) or numerical simulations (Bakun-Mazor et al., 2009; Cała et al., 2016; Kortnik, 2015; Lafrance et al., 2016; Oggeri \& Oreste, 2015; Parise \& Lollino, 2011; Tritsch et al., 2007). Cave SHM may be achieved via regular visual inspections, in-situ geotechnical monitoring (displacement sensors, pore-pressure, water level loggers), teledetection surveys (photogrammetric or laser measurements) and geophysical measurements (seismic and acoustic crack detection, e.g. Hanesse and Richard (2016), Xu et al. (2015), Idrees and Pradhan (2018), Bartolo and Salvini (2019)). Although accurate and informative, such monitoring surveys remain generally sparse both in space and time, assuming homogeneous internal properties of the rock mass (Fargier et al., 2017), with often arbitrary approximations.

Geophysical techniques have proven to be useful for risk mitigation studies as they may reveal hidden geological structures and yield physical parameters of rock pillars and burdens despite the complex 2D or 3D geometry of such targets (Das et al., 2017; Lesparre et al., 2014; Marescot et al., 2008). Most studies exploit electrical resistivity tomography (ERT), refraction seismic tomography (RST) and/or ground penetrating radar (GPR) e.g. Leucci et al. (2007), Cardarelli et al. (2010), Cote et al. (2005), Fargier et al. (2017), Dérobert and Abraham (2000), Jones et al. (2014), Li et al. (2015), Srivastava et al. (2020) and Kumar et al. (2021).

Electrical resistivity $\rho$ is primarily controlled by porosity, water conductivity and water content (Archie et al., 1942; Brunet et al., 2010) while seismic pressure-wave velocity $V_P$ depends on water content, rock stiffness, density and porosity (Coutant et al., 2012; Draebing \& Krautblatter, 2012; Hauck et al., 2011; Timur, 1968). Both geophysical techniques suffer from intrinsic ambiguities, failures and non-uniqueness issues (Doetsch et al., 2010; Gallardo \& Meju, 2004; Maurer \& Hauck, 2007; Lelièvre et al., 2012; Linder et al., 2010). Thus, the combined use of ERT and RST represents a promising tool, which has been recently applied to many targets (Carrier et al., 2019; Díaz et al., 2014; Ganerød et al., 2006; Linder et al., 2010; Rønning et al., 2014). The combination of geophysical techniques may be achieved with many processing flows (Coutant et al., 2012; Doetsch et al., 2012; Gallardo \& Meju, 2004; Gunther \& Rucker, 2006; Gallardo \& Meju, 2011; Hellman et al., 2017; Ronczka et al., 2017; Wagner et al., 2019). These processes can be classified into several categories, notably Independent Inversions with joint interpretation (IIv) and Joint Inversion (JIv) (Gallardo and Meju 2004; Hellman et al. 2017; Lelièvre et al., 2012). Independent Inversion with joint interpretation (IIv) consists in the individual processing of each geophysical method followed by a common model merging (Bedrosian et al., 2007; Cardarelli et al., 2010; Eberhart-Phillips et al., 1995; Hirsch et al., 2008; Tronicke et al., 2004). The separate inversion processes conducted in IIv may yet lead to diverging, inconsistent results between methods which are analyzed and inverted independently (Mewes et al., 2017). The joint inversion (JIv) relies on method coupling during the inversion process to improve model resolution, consistency and decrease model ambiguities (Gallardo \& Meju, 2004; Gunther \& Rucker, 2006; Hellman et al., 2017; Lelièvre et al., 2012; Mewes et al., 2017; Ronczka et al., 2017).

Petrophysical joint inversion (P-JIv) consists in using pre-existing petrophysical laws relating the inverted physical parameters (Linde et al., 2006; Wagner et al., 2019). These laws generally derive from empirical laboratory experiments conducted on small-scale samples and generalized in the field (Linde et al., 2006; Pellet et al., 2016; Wagner et al., 2019). P-JIv might hence suffer from sample non-representativity issues, especially if the electrical properties of the investigated medium vary across the site and/or with time. Method coupling may also be achieved by enhancing common features between geophysical models during the inversion process. This so-called structural joint inversion (S-JIv) was first introduced for ERT and RST by Gallardo and Meju (2003) and Gallardo and Meju (2004). In porous media, P-wave seismic velocity $V_P$ and electrical resistivity $\rho$ are generally well correlated because both parameters depend on porosity (Archie et al., 1942; Meju et al., 2003; Wyllie et al., 1956). Gallardo and Meju (2004) hence proposed a cross-gradient approach where a penalty factor applies if model gradients do not follow the same direction. Ronczka et al. (2017) and Hellman et al. (2017) extended structural-constrained joint-inversion to non-regular grids and reciprocal constraints between methods via the adaptation of the constraint matrix at every iteration.

Whatever the inversion type, a good estimation of the error on the data appears as a key parameter to drive the process (Tarantola, 2005; Tso et al., 2017). It is notably essential for computing the objective function and avoiding data over- or under-fitting. The total data error corresponds to the sum of measurement uncertainties and data error weight considered for the inversion. In addition, the computation of the resolution is crucial to prove the reliability of the model (Tarantola, 2005).

In this paper, we present the results of a geophysical survey conducted in an abandoned quarry. The aim is to investigate the structural health of limestone pillars. Considering the sensitivity of the collected data, the manager of the studied quarry has chosen to remain anonymous. As a consequence, some technical reports used as references will also remain confidential. Both IIv and P-JIv processing of ERT and RST datasets are presented in the manuscript. S-JIv was also tested but is not shown here as it yielded similar results as P-JIv. For P-JIv, we first define a simplified petrophysical 3 phases model derived from Archie's and Wyllie's laws to link electrical resistivity and P-wave seismic velocity. We then apply a clustering algorithm in the seismic velocity-electrical resistivity space to identify areas with similar physical properties. This aims at classifying the structural health of limestone pillars with a quantitative, unbiased indicator. Please note that the aim of the paper is not much to develop the inversion method but rather to show its applicability and usefulness for specific targets, i.e. estimate damage degree of limestone pillars in underground quarries.

\section{Geological Settings}

\begin{figure}[htbp]
\centering
\includegraphics[width=0.55\textwidth]{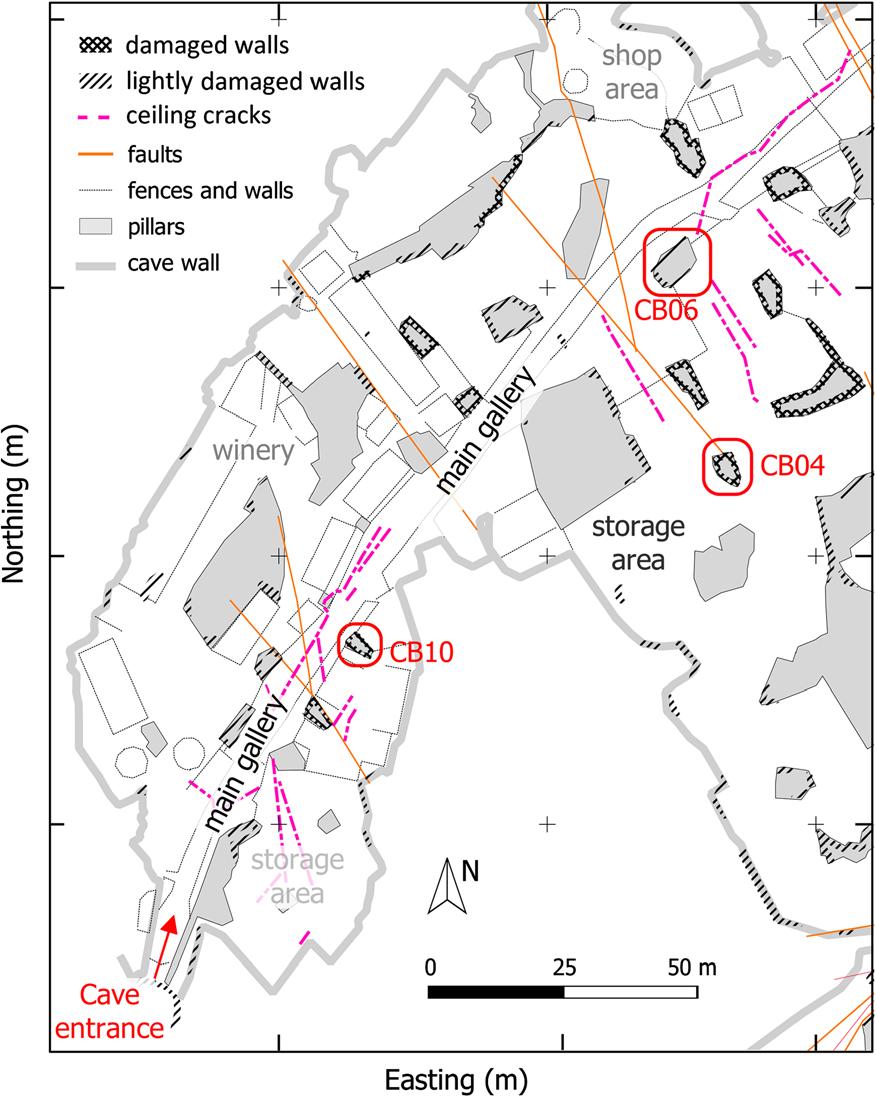}
\caption{Map of the abandoned quarry (chamber-pillar). Precise geographic coordinates are voluntarily hidden.}
\label{fig:map}
\end{figure}

The studied winery plant is located in an abandoned underground quarry in Burgundy, France (see location in Fig.~\ref{fig:map}). The winery represents a major economical and touristic player in Burgundy. About 3 million bottles of Cremant wine are produced every year with the grapes brought by about 430 winegrowers. The quarry consists of a set of chambers and pillars dug in sub-horizontal, about 18--25~m thick Kimmeridgian oolitic chalky limestones covered by Middle to Upper Kimmeridgian marls. Quarry excavation began during the 5th century AD and progressively spread up to about 4 hectares at present (Fig.~\ref{fig:map}). Detailed geotechnical surveys conducted by the French National Geohazard service (technical reports, confidential) and SAGE consulting company showed that the general state of the cave was satisfactory, although some areas required specific attention. These areas are affected by clear geological fractures (N0, N100 and N140 fracture sets, Fig.~\ref{fig:map}). Cracking processes were detected at cave roof (limestone strata delamination) especially near the cave entrance and in the central storage area (ceiling cracks, purple dashed lines in Fig.~\ref{fig:map}). Specific areas on rock walls and pillars showed slightly damaged areas (e.g. wall peeling shown with hatches in Fig.~\ref{fig:map}) or more severely damage areas (e.g. wall fracturing shown with cross-hatches in Fig.~\ref{fig:map}). A numerical stability study revealed that the central storage area was subject to the most vertical stresses due to small area of the pillar and their sparsity (technical reports, confidential). Laboratory analyses conducted by the French National Geohazard service on broken rock samples revealed low compressive strength, lying in the range $9.8 < R_c < 14.8$~MPa. In comparison, values of $20 < R_c < 40$~MPa are commonly measured for similar quarry stones. Young modulus $E$ ranged between 3.6 and 4.9~GPa. Dry sample density ranged about 2.0. The rock strength appears highly variable depending on its water content. It notably reached a pasty state during drilling, when saturated. Water dripping occurs from the cave roof, with higher intensity near the entrance where the rock burden is thinner. Given this observation, the winery owners mandated the SAGE company in 2019 in order to design and prioritize the reinforcement works. Three pillars were selected for a detailed geotechnical and geophysical survey because of the need in-depth information to adapt their reinforcement design: CB10, CB04 and CB06 (circled in red in Fig.~\ref{fig:map}). The CB10 pillar showed a significant damage degree with intense rock peeling and internal destructuration caused by several subvertical fractures. Its North, East and South corners appeared particularly damaged. CB04 pillar showed damage, with surface peeling on all the faces and diabolo-shape. CB06 revealed limited damage, with surface peeling on some areas on its North-West and South-West faces and open cracks at its western corner.

\section{Data Acquisition and Pre-processing}

\subsection{Drilling Logs}

\begin{figure}[htbp]
\centering
\includegraphics[width=\textwidth]{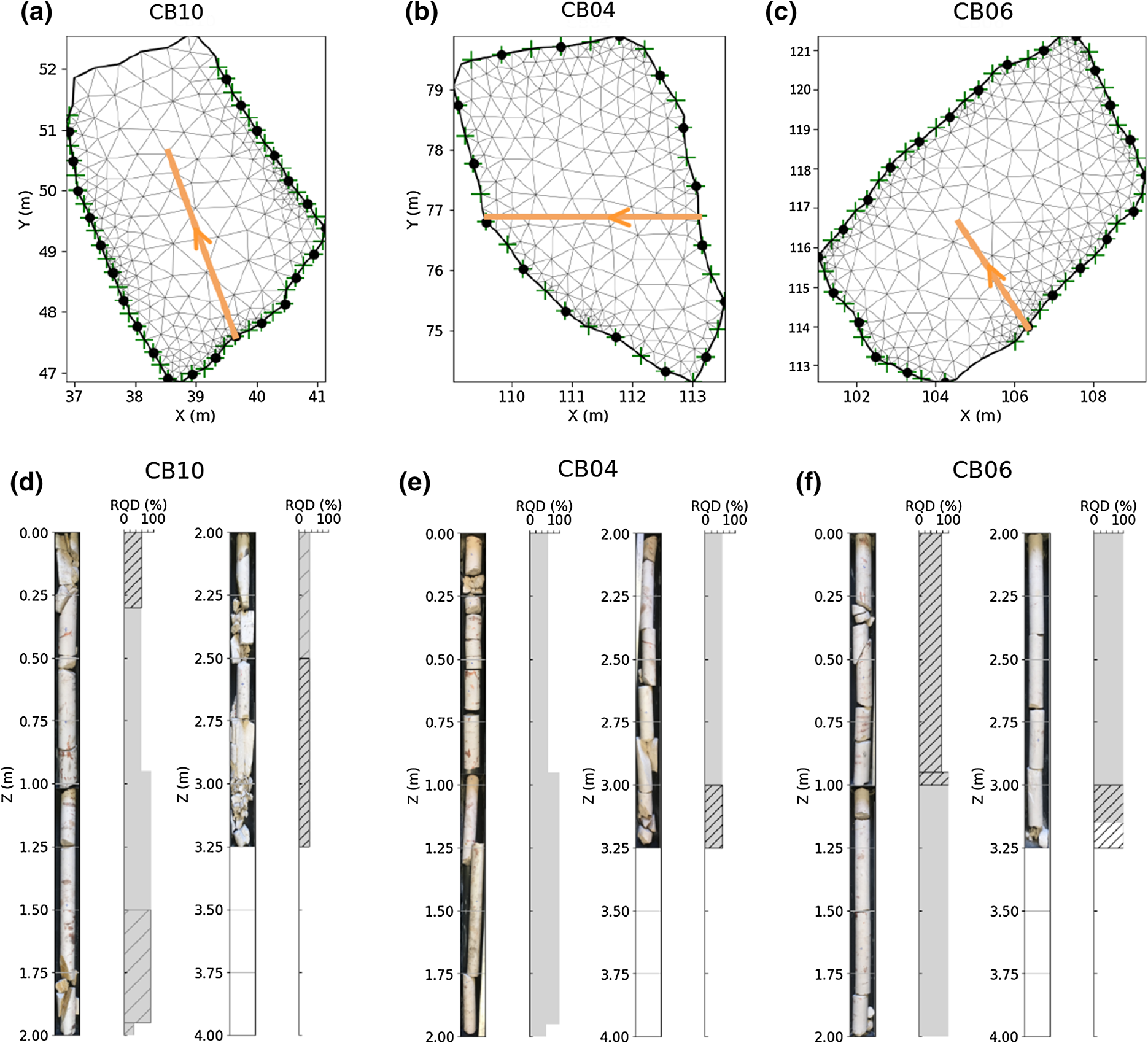}
\caption{a--c Location of ERT sensors (black circles), RST geophones (green crosses) and drilling holes (orange lines) for CB10, CB04 and CB06 pillars, respectively. Arrows indicate the direction of drilling. d--f Pictures of the drilling cores are shown on the left while corresponding Rock Quality Designation RQD is shown at right. Black striped areas: observed fractures on core samples. Grey striped areas: moderately fractured samples with oxidation traces (orange deposits). a--c Are not drawn at the same scale due to different pillar sizes.}
\label{fig:setup}
\end{figure}

A 68~mm diameter, single envelope core drill was used in each of the three pillars, using water as drilling fluid. 3.25~m long boreholes were bored. Drilling core locations are shown in Fig.~\ref{fig:setup}a--c. Sampling recovery percentage and Rock Quality Designation were computed for the entire logs. Porosity, water content, uniaxial compressive strength ($R_c$) and ultrasonic velocity tests were carried out on core samples. Some samples were saturated with water during a few days to a few weeks prior to the $R_c$ tests in order to better simulate on-site conditions.

Pictures of drilling logs performed on CB10 (left), CB06 (center) and CB04 (right) are shown in Fig.~\ref{fig:setup} at the bottom. The cores showed a very good sampling recovery ratio (generally 100\% except 70\% for last meter of CB10). Rock Quality Designation (RQD) ranged between 34 and 100\% (grey bars, Fig.~\ref{fig:setup}). Whatever the pillar, the RQD being generally higher close to pillar core than close to surface. Indeed, CB10 showed very damaged areas from 1.7 to 3.2~m as well as in the first 20~cm (black hatched areas, Fig.~\ref{fig:setup}). CB04 drilling bored the pillar from one side to the other. We note that the center of the core is less damaged than the borders, the first and the last meters showing lower RQD. Similarly, the quality of the rock increases from the border to the center for CB06. Uniaxial compression tests yielded $R_c$ values ranging from 16.2 to 36.9~MPa with mean of 26~MPa and standard deviation of 7~MPa. Higher $R_c$ are observed on rock samples taken from core sections with higher RQD, confirming that pillar cores have better mechanical properties than the external fractured zone. These $R_c$ values are about twice higher than former lab tests, probably because the latter were conducted on broken rock samples (see Sect.~2).

Water content and porosity laboratory measurements were conducted on good quality samples. The connected porosity ranges between 30 and 36\% and is equal to the total porosity which is consistent with the values known for chalk or chalky limestones (Dessandier et al., 2006; Fabricius, 2020). The water content ranges from 15 to 20\% (the highest water content and porosity being obtained for CB04, the lowest water content and porosity being obtained on CB06).

\subsection{Electrical Data}

ERT sections were acquired horizontally around the pillars at about 1~m above the cave floor. Stainless steel electrodes were placed into hammer drill holes filled with bentonite to ensure good contact resistances. In order to fit with the perimeter of the pillars, 1~m spacing between the electrodes was used for the pillars CB04 and CB06 and 0.5~m spacing for CB10. Electrode coordinates were measured on a high-resolution lidar digital surface model (DSM). Electrodes location is shown on Fig.~\ref{fig:setup}. A multichannel ABEM Terrameter SAS4000 was used in combination with an electrode selector ES1064 (Lund Imaging system). A Wenner--Schlumberger scheme was performed on CB04, CB06 and CB10, respectively resulting in 71, 183 and 187 data acquisition measurements.

\begin{figure}[htbp]
\centering
\includegraphics[width=0.5\textwidth]{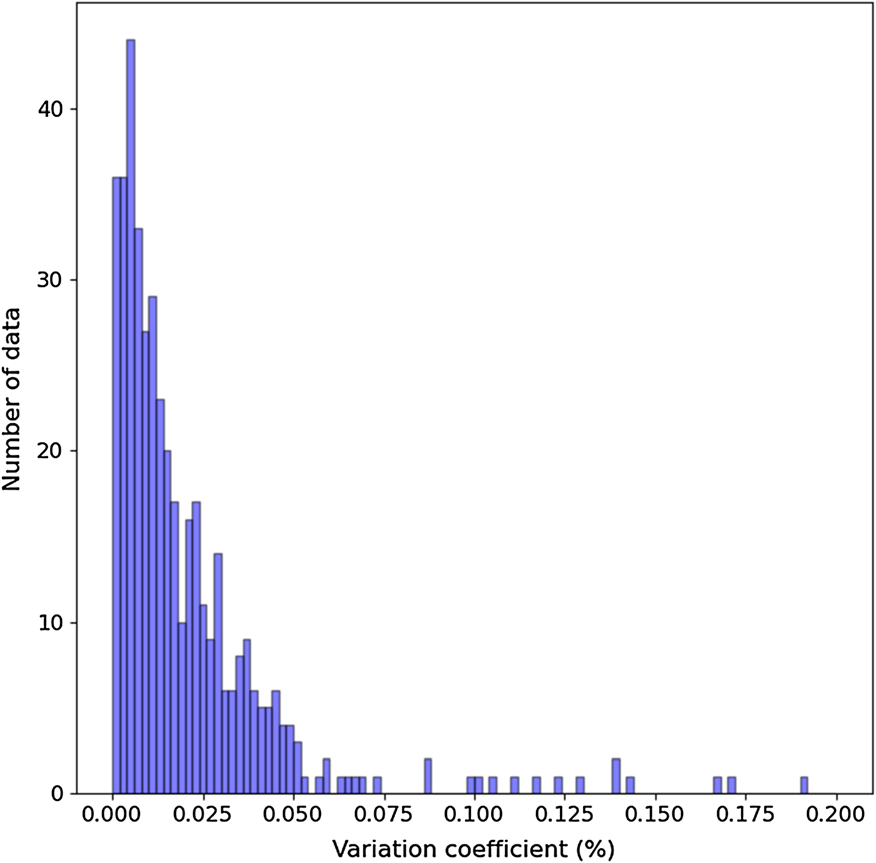}
\caption{Histogram of ERT measurement stacking error---also called variation coefficient---for the 3 pillars CB10, CB04 and CB06 together. For readability, values ranges from 0 to 0.2\%, skipping the 10 higher values (between 0.2 and 3\%).}
\label{fig:stack}
\end{figure}

\begin{figure}[htbp]
\centering
\includegraphics[width=\textwidth]{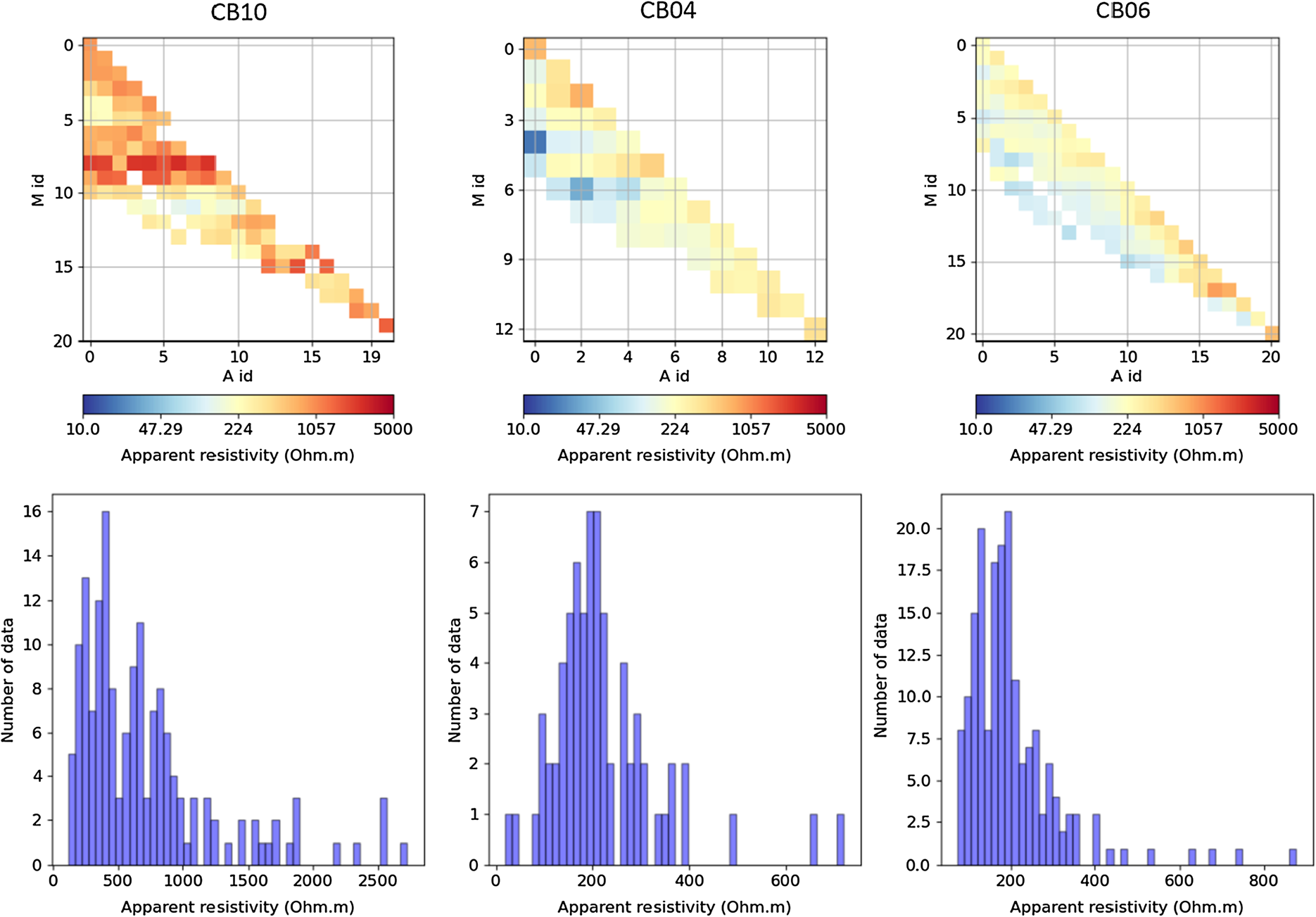}
\caption{(Top row) electrical data after outlier removal for CB10, CB04 and CB06 pillars, respectively. Due to the non-conventional geometry of acquisition, apparent resistivities are arbitrarily represented as a function of A-injection and M-reception electrodes position. (Bottom row) distribution of apparent resistivities is presented for the same pillars. All three subplots do not share the same spatial scale.}
\label{fig:ertdata}
\end{figure}

Injected current ranged between 20 and 100~mA. We measured the stacking errors (see definition in Tso et al. (2017) and Parsekian et al. (2017)) over a minimum of 3 injection cycles. These errors---also called variation coefficients---are shown in Fig.~\ref{fig:stack}. They ranged between 0.0004 and 3\%, with a mean and median of 0.04 and 0.01\% (0.18 and 0.1~mV) respectively. These values suggest a very good signal/noise ratio. We did not realize reciprocal measurements, nor contact resistances measurements. Apparent resistivities were computed from measured resistances using geometrical factors deriving from high-resolution lidar DSM. Geometrical factors were computed using the tools from pyGIMLI toolbox (Günther et al., 2006; Rücker et al., 2006, 2017). Based on the distribution of the apparent resistivities, 20, 1 and 0 outliers (apparent resistivity values that differs significantly from the others, here $>3000$ or $<3$~Ohm~m) have been removed for CB10, CB06 and CB04, respectively. The resulting dataset for each pillar is shown as a function of the injection electrodes position (A, B) and the receivers electrodes position (M, N) in Fig.~\ref{fig:ertdata}. Table~\ref{tab:ert} shows the data statistics and filtering.

\begin{table}[htbp]
\centering
\caption{Data statistics and filtering for ERT}
\label{tab:ert}
\begin{tabular}{lllllllll}
\toprule
Pillar & \multicolumn{7}{l}{Data, resistivities (Ohm m)} \\
\cmidrule(l){2-8}
 & Removed & Initial nb & Final nb & Min & Max & Mean & Std \\
\midrule
CB10 & E15 and $\rho>3000$ & 187 & 167 & 106 & 1944 & 619 & 413 \\
CB06 & $\rho<3$ & 183 & 182 & 73 & 876 & 202 & 113 \\
CB04 & / & 71 & 71 & 19 & 720 & 220 & 115 \\
\bottomrule
\end{tabular}

\smallskip
\parbox{0.85\textwidth}{\footnotesize Data are expressed as apparent resistivities in Ohm m. Minimum, maximum, mean values and standard deviation are expressed after filtering on the final number of data. \emph{nb} number of data.}
\end{table}

\subsection{Seismic Refraction Data}

RST spreads were located at the same location as ERT profiles. 14~Hz horizontal geophones were placed normally to the pillar rock face into hammer drill holes and set with plaster. 1~m spacing was used for CB04 and CB06 and 0.5~m spacing for CB10. Shots have been performed at mid-distance between consecutive geophone positions. Sensors and shots locations are shown in Fig.~\ref{fig:setup}. We used a hammer source with hammer-switch trigger of 0.1~ms precision. Signals were recorded by a Geometrics Geode at 4000~Hz sampling frequency. As stated in Cote et al. (2005), seismic tomography on small-scale targets with relatively fast velocities reveals challenging in terms of instrumentation. Arrival time picking was carried out carefully as traveltimes range about a few milliseconds only.

\begin{figure}[htbp]
\centering
\includegraphics[width=\textwidth]{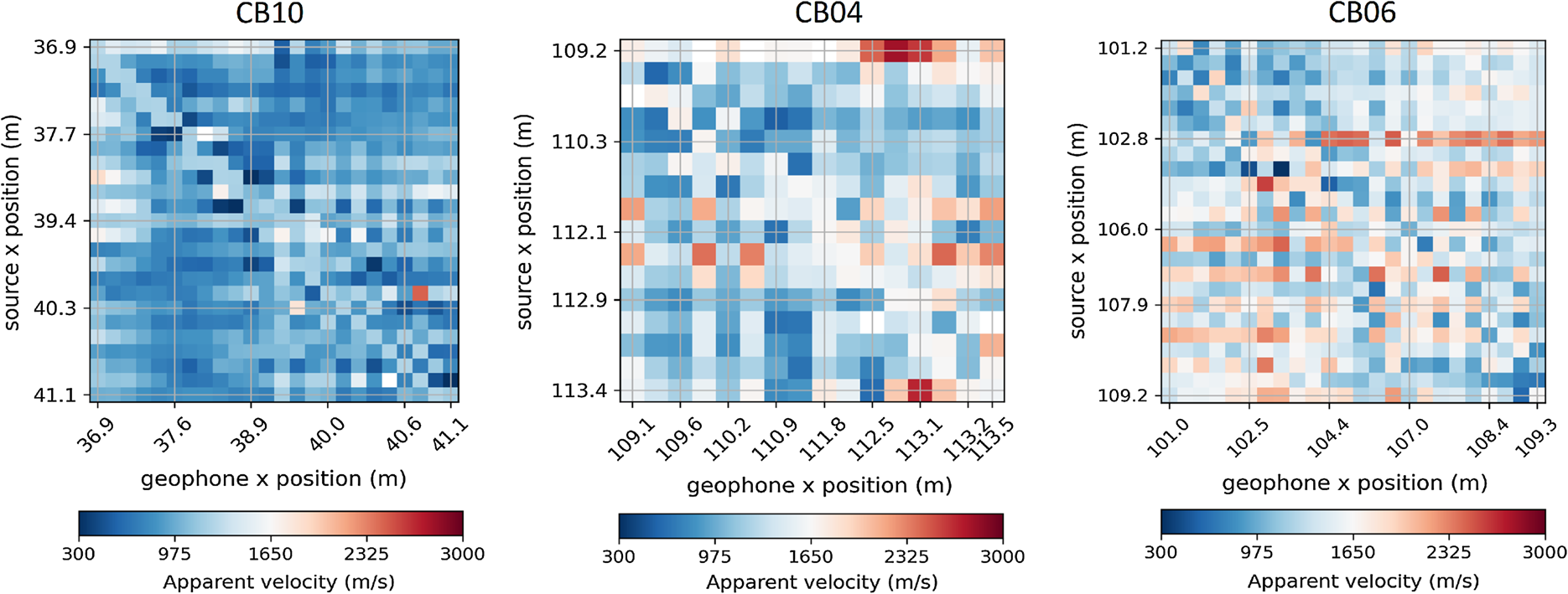}
\caption{Seismic data after outlier removal for CB10, CB04 and CB06 pillars, respectively. Apparent velocities are arbitrarily represented as a function of geophones and source horizontal coordinates $x$ in meters. Due to this peculiar geometrical arrangement, apparent velocities are not expected to show a smooth pattern. All three subplots do not share the same spatial scale.}
\label{fig:rstdata}
\end{figure}

Apparent velocities distribution has been checked in order to identify potential outliers, eliminating doubtful velocities with $V>4000$ or $V<300$~m/s. The higher boundary corresponds to the maximum P-wave seismic velocity expected in the studied limestones. It derives from the observed porosity and laboratory measurements on intact samples showing $V_P<3500$~m/s. The lower boundary roughly corresponds to P-wave seismic velocity in the air. Outliers were removed based on the distribution of apparent velocities for each pillar. Apparent velocities after filtering are shown in Fig.~\ref{fig:rstdata}. Data quality analysis is shown in Table~\ref{tab:rst}.

\begin{table}[htbp]
\centering
\caption{Data statistics and filtering for RST}
\label{tab:rst}
\begin{tabular}{lcccccc}
\toprule
 & \multicolumn{6}{c}{Data, apparent velocities $V$ (m/s)} \\
\cmidrule(l){2-7}
 & Initial nb & Final nb & Min & Max & Mean & Std \\
\midrule
CB10 & 600 & 597 & 306 & 2446 & 948 & 265 \\
CB06 & 576 & 576 & 318 & 2575 & 1475 & 388 \\
CB04 & 256 & 250 & 544 & 2781 & 1348 & 421 \\
\bottomrule
\end{tabular}

\smallskip
\parbox{0.75\textwidth}{\footnotesize Data are expressed in apparent velocities $V$ in m/s. Minimum, maximum, mean values and standard deviation are expressed after filtering on the final number of data. \emph{nb} number of data.}
\end{table}

\section{Inversion of Geophysical Data}

\subsection{Meshing}

Refraction seismic tomography (RST) and electrical resistivity tomography (ERT) have been performed using boundless electrical resistivity tomography (BERT) and pyGIMLI tools developed by Günther et al. (2006), Rücker et al. (2006) and Rücker et al. (2017). A unique irregular closed mesh is built considering electrodes and geophones positions, as well as the topography of each pillar. The same grid has been used to perform individual (IIv) and joint (P-JIv) inversions for each individual pillar.

\subsection{Independent Inversions (IIv)}

In independent inversions, $\bm{d}$ is a $n$-dimension vector of the observed data and $\bm{m}$ the $p$-dimension vector of the model parameters. Knowing $f(\bm{m}) = \bm{d}$ the forward modelling operator mapping the physical relationship between the model parameters and the data, the inversion process consists in minimizing the following objective function:
\begin{equation}
\Phi = \Phi_d + \lambda \Phi_m
\label{eq:obj}
\end{equation}
where $\Phi_d = \| \bm{W_d}(f(\bm{m}) - \bm{d}) \|_2^2$, $\Phi_m = \| \bm{W_m}(\bm{m} - \bm{m_0}) \|_2^2$ and $\lambda$ is the regularization parameter determined via a L-curve process (Günther et al. 2006). $\bm{W_d}$ is the diagonal data weighting matrix containing the inverse of the error estimate on the data, $\bm{W_m}$ is a constrained matrix acting as a roughness operator and $\bm{m_0}$ is a reference model. In the following $\bm{m_0}$ is considered as a constant for each inversion. $\bm{W_m}$ is the derivative matrix applied to $\bm{m}$ which controls the model characteristics (Günther et al. 2006). We assumed a Gaussian data error distribution whose standard deviation is the sum of measurement and modelisation standard deviations (Tarantola 2005; Monteiller 2005). In BERT and pyGIMLI software, the error estimate on the data is considered as the sum of the measurement accuracy and a fixed percentage (Günther et al. 2006; Lajaunie et al. 2019). For ERT and RST, we set the fixed percentage of error to 3\% and 1\%, respectively. The measurements errors were set to 0.1~mV and 0.1~ms, respectively, considering the accuracy of the measurements on the field. A L2-norm is considered for $\Phi_m$ as relatively smooth contrasts are expected in the model. If some significant outliers are observed in the data misfit distribution, $\Phi_d$ can be expressed as a L1-norm.

The minimization of the objective function is performed on the basis of a generalized Gauss--Newton algorithm (Rücker et al. 2017). At the $k$th step, the expression of the model is then:
\begin{equation}
\begin{split}
\left(\bm{J}^T \bm{W_d}^T \bm{W_d} \bm{J} + \lambda \bm{W_m}^T \bm{W_m}\right) \Delta \bm{m}^k
= \bm{J}^T \bm{W_d}^T \bm{W_d} \left(\bm{d} - f(\bm{m})\right) - \lambda \bm{W_m}^T \bm{W_m} \left(\bm{m}^k - \bm{m_0}\right)
\end{split}
\label{eq:gn}
\end{equation}
where $\Delta \bm{m}^k = \bm{m}^k - \bm{m}^{k-1}$ and $\bm{J} = \frac{\partial f(\bm{m})}{\partial \bm{m}}$ is the Jacobian matrix.

Equation~(\ref{eq:gn}) is solved using a conjugate-gradient least-squares solver (Günther et al. 2006). The stopping criterion is reached when either the cost function decreases by less than 2\% or $\Phi_d$ satisfies the $\chi^2$ criterion (i.e. $\frac{\Phi_d}{n}$ is close to 1). If none of these criteria are reached, a reasonable maximum of 5 iterations was set.

For ERT, the reference model is used as a starting model. It is considered as an homogeneous model with the median of the measured apparent resistivities. For RST, the starting model is a homogeneous P-wave velocity model where $V_P = 1500$~m/s. The coverage is defined as the sum of the absolute sensitivities (i.e. Jacobian matrix) for ERT and through the density of rays for RST.

\subsection{Petrophysical Joint Inversion (P-JIv)}
\label{sec:pjiv}

Electrical bulk conductivity can be expressed as a function of porosity, grain size and tortuosity, fluid saturation, fluid electrical conductivity and surface conductivity (Revil, 2013; Romero-Ruiz et al., 2018). Surface conductivity is a function of cation exchange capacity or specific surface area. It cannot be neglected in shaly and clayey sediments (Revil et al., 1998; Waxman et al., 1968). However, some authors suggested that the surface conductivity cannot be neglected even in non-clayish rocks, if the water in the pores is fresh enough. This is the case when the water conductivity is small, i.e when the salinity remains negligible (Bolève et al., 2007; Leroy et al., 2008; Revil \& Linde, 2006; Revil et al., 1999). In Duvillard et al. (2018), the low salinity limit corresponds to pore water conductivity less than 0.03~S/m (pore water resistivity more than 30~Ohm~m). In our study area, the water in the pores is drinkable. The technical report published in 2015 by the Geological Office of France (BRGM, 2015) also suggests that the water contained in the Jurassic rocks in our area does not present large chemical heterogeneities. It shows resistivities between 15 and 25~Ohm~m. We considered the relationship between surface conductivity, pore water conductivity and bulk conductivity established by Revil and Linde (2006) and further used by Linde et al. (2006). For petrophysical joint inversion in the high salinity range with a pore water conductivity of 0.05~S/m and a porosity of 30\% and water saturation of 100\% (values obtained for our samples at the core), a bulk conductivity at the core of 100~Ohm~m (retrieved from independent inversion), the surface conductivity in our case would be the order of 0.006~S/m which is one order of magnitude less than the pore water conductivity. Furthermore, as no clay is observed in the limestones in the cave, we reasonably assumed a non-conductive rock medium. This implies that surface conduction is negligible so that the electrolytic conduction dominates. We performed the joint inversion (P-JIv) considering a 3 phase petrophysical model (3PM). This model was written based on Hauck et al. (2011) and Wagner et al. (2019) 4 Phase Model (4PM). The relation between resistivity, porosity and saturation was given by Archie's law (Archie et al., 1942).
\begin{equation}
\rho = a \rho_w \Phi^{-m} S_w^{-n}
\label{eq:archie}
\end{equation}
where $a$ is a constant usually close to 1, $\rho_w$ is the resistivity of the liquid in the pores, $\Phi$ is the porosity of the rock defined as the volume of the connected pores divided by the total volume, $S_w$ is the liquid saturation and $m$ and $n$ are the cementation and saturation exponents. The resistivity becomes a function of the porosity and liquid (water) content.

Archie's law is applicable in the case of porous rocks with a non-conductive matrix filled or partially filled with brine. As it does not consider surface conduction, it especially fails when clay is present. It also fails if water saturation or porosity are too low (Glover, 2010; Tsallis et al., 1992; Worthington, 2005). Here, surface conduction has been demonstrated as being negligible. As laboratory experiments showed that water saturation is high, the total porosity equals the effective porosity (i.e. all the pores are connected). The porosity of the considered chalky limestones is about 30\%, which assures us to be in the applicability domain of Eq.~(\ref{eq:archie}).

For seismic P-wave velocity, Wyllie et al. (1956) and Timur (1968) have shown that the slowness of a medium can be express as the sum of the slownesses of its individual components weighted by their volumetric fractions. This assumption is valid for isotropic media with single homogeneous mineralogy and under sufficient effective pressure (Hauck et al., 2011; Wagner et al., 2019).

The formulation given in the 4PM by Hauck et al. (2011) becomes the following for 3PM:
\begin{equation}
\frac{1}{v} = \frac{f_r}{v_r} + \frac{f_w}{v_w} + \frac{f_a}{v_a}
\label{eq:wyllie}
\end{equation}
where $f_r, f_w, f_a$ refer respectively to the rock fraction, the water fraction and the air fraction of the medium such as $f_r + f_w + f_a = 1$, $0 \le f_r, f_w, f_a \le 1$, and $v$, $v_r$, $v_w$, $v_a$ refer to the mean velocity, the velocity in rock, water and air respectively.

The porosity being defined as $\Phi = 1 - f_r$ and the saturation as $S_w = \frac{f_w}{\Phi}$, Eqs.~(\ref{eq:archie}) and (\ref{eq:wyllie}) enable to formulate water and air content as a function of rock content:
\begin{equation}
f_w = \left( \frac{a \rho_w (1-f_r)^n}{\rho (1-f_r)^m} \right)^{1/n}
\label{eq:fw}
\end{equation}
\begin{equation}
f_a = v_a \left( \frac{1}{v} - \frac{f_r}{v_r} - \left( \frac{a \rho_w (1-f_r)^n}{\rho (1-f_r)^m} \right)^{1/n} \frac{1}{v_w} \right)
\label{eq:fa}
\end{equation}
The joint inversion consists in adding an additional vector parameter $\bm{p}$ representing the volumetric fraction of rock, air and water according to Wagner et al. (2019) model. The data vector is composed of a concatenation of observed traveltimes and logarithmized apparent resistivities. Logarithmic barriers are used to keep the volumetric fractions between 0 and 1. The new objective function to minimize takes into account additional regularization terms to constrain volumetric fractions:
\begin{equation}
\begin{split}
\Phi_{joint} = {} & \| \bm{W_d}(f(\bm{m}) - \bm{d}) \|_2^2 + \alpha^2 \| \bm{W_m}(\bm{m}) \|_2^2 \\
& + \beta^2 \| \bm{W_p^{sum}}(\bm{p} - \bm{1}) \|_2^2 + \gamma^2 \| \bm{W_p}(\bm{p} - \bm{p_0}) \|_2^2
\end{split}
\label{eq:joint}
\end{equation}
The variables and terms are the very same as the ones used in Wagner et al. (2019). The third and fourth term in Eq.~(\ref{eq:joint}) ensure respectively volume conservation and damping regularization penalizing solutions deviating from a reference model $\bm{p_0}$. $\bm{W_p^{sum}}$ is a block matrix enabling the sum of the volumetric fractions to stay close to 1 and $\bm{W_p}$ is a square matrix enabling to select the model parameters that should stay close to the reference model. $\alpha$, $\beta$ and $\gamma$ are regularisation parameters. A least-square minimization of Eq.~(\ref{eq:joint}) is performed using a LSQR algorithm.

Considering laboratory data, the velocities of rock, air and water were fixed at 4000, 300 and 1500~m/s respectively. The initial porosity was fixed to 0.2 as reasonable a priori guess and the resistivity of water was set to 20~Ohm~m (BRGM, 2015). No laboratory experiment were available from rock samples of our survey area to determine $m$ and $n$ exponents in Eq.~(\ref{eq:archie}). We fixed $m = 1.4$ and $n = 2.4$ as in Wagner et al. (2019). These values agree with most of previous studies (Coperey et al., 2019; Duvillard et al., 2018; Hermans et al., 2012; Revil, 2013; Revil et al., 1998) and yielded realistic results.

The total air fraction is the sum of the air fraction in the pores, $f_{a(pores)}$, and the air fraction in the fractures, $f_{a(fractures)}$. The total water fraction is the sum of the water in the pores, $f_{w(pores)}$, and the water in the fractures, $f_{w(fractures)}$. Considering that the pores are water saturated, $f_{a(pores)} = 0$, the fraction of fractures consists in the sum of the air fraction and the water fraction in the fractures: $f_{fractures} = f_{a(fractures)} + f_{w(fractures)} \sim f_a + f_{w(fractures)}$. This suggests that in the case of opened and drained fractures ($f_{w(fractures)} \sim 0$), the air fraction can be used as a proxy of fractures fraction and at least as a minimum estimation of fractures fraction. However, no measurements of $m$ and $n$ parameters of the Archie's law are available in our case. Thus, we will not use laboratory measures of the porosity to help the interpretation of the obtained models. We neither try to interpret quantitatively our results. Please note that the main point of this paper is to highlight the feasibility and potential of the method for this given context and not to discuss about laboratory data or petrophysical parameters.

\subsection{Model Coverage}

We used the alternative approach to the resolution by computing the cumulative sensitivity map (Günther et al., 2006). For ERT, the coverage is defined as the sum of the absolute values of the sensitivity matrix (i.e. Jacobian) for each model parameter. The logarithm of the Jacobian matrix is used to get rid of geometrical effects (\ref{eq:G}). The coverage is then expressed as in Eq.~(\ref{eq:cov}).
\begin{equation}
G_{ij} = \frac{\partial \log\left(f(m)_i\right)}{\partial \log(m_j)}
\label{eq:G}
\end{equation}
\begin{equation}
Coverage_j = \sum_{i=1}^{n} \frac{G_{ij}}{cellsize_j}
\label{eq:cov}
\end{equation}
The RST coverage was set to 1 if a ray crosses a given cell, 0 in the other case. For joint inversions, we multiply the RST coverage to the ERT coverage to obtain a combined coverage representing the degree of information brought by the two models.

To get additional information about model appraisal, and because this quantity can differ from the coverage, we choose to compute the depth of investigation (DOI) index for ERT (Caterina et al., 2013; Hilbich et al., 2009; Oldenburg \& Li, 1999; Oldenborger et al., 2007). Initially the DOI index has been established in order to find the depth under which the surface data becomes insensitive to the physical properties of the medium (Oldenburg \& Li, 1999). Let $m_0^1$ and $m_0^2$ be two starting models such as $m_0^1 = 10 \times m_0^2$ and $\bm{m}^1$ and $\bm{m}^2$ the inversion resulting models associated. The DOI index is defined as (Marescot et al., 2003; Oldenburg \& Li, 1999):
\begin{equation}
DOI = \frac{\bm{m}^1 - \bm{m}^2}{m_0^1 - m_0^2}
\label{eq:doi}
\end{equation}
If the DOI is close to 0, then the two inversions produce the same results independently from the starting model. This means that model information in these areas have been brought by the data. In contrast, when the DOI is close to 1, the inversion result depends on the starting model and few or no information has been added by the data. We apply the DOI computation for our P-JIv model later in the article.

\subsection{Clustering Method}
\label{sec:clust}

Pillar geophysical cross-sections are then processed through a clustering algorithm which aims at identifying consistent (electrical resistivity, P-wave velocity) areas. We used a Mean Shift clustering algorithm (Pedregosa et al., 2011) because of its good performance in handling the complexity of real data sets (Ronczka et al., 2017), contrary to other clustering techniques (Comaniciu \& Meer, 2002; Jain et al., 2000). The feature space is composed of (electrical resistivity, P-wave velocity) couples. The bandwidth controlling the analysis of the feature space is automatically calculated from the data points and a predefined quantile. The quantile controls the pairwise distance used and hence the final number of clusters. For cluster computation, only the data with a non-zero coverage (i.e. non-zero ray coverage) are used.

\section{Results}

\subsection{Geophysical Prospecting}

\begin{figure}[htbp]
\centering
\includegraphics[width=\textwidth]{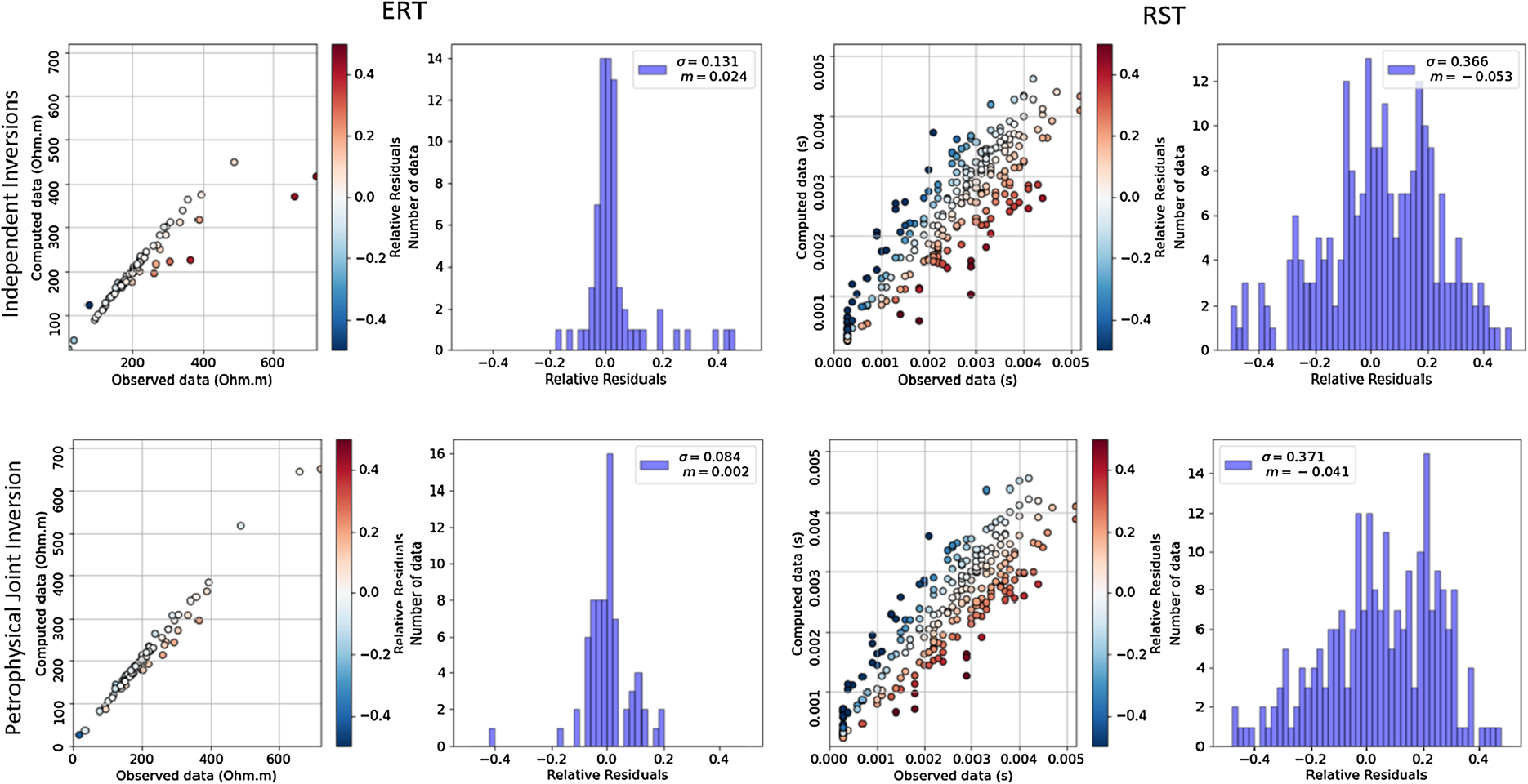}
\caption{CB04 pillar. Cross-plots (scatter plot, left) and residuals (histogram, right) are shown both for ERT (left column) and RST data (right column). First rows shows the Independent Inversion (IIv) and bottom row shows the Petrophysical Joint Inversion (P-JIv). The color scale of scatter points represents the relative residuals.}
\label{fig:qc04}
\end{figure}

\begin{figure}[htbp]
\centering
\includegraphics[width=\textwidth]{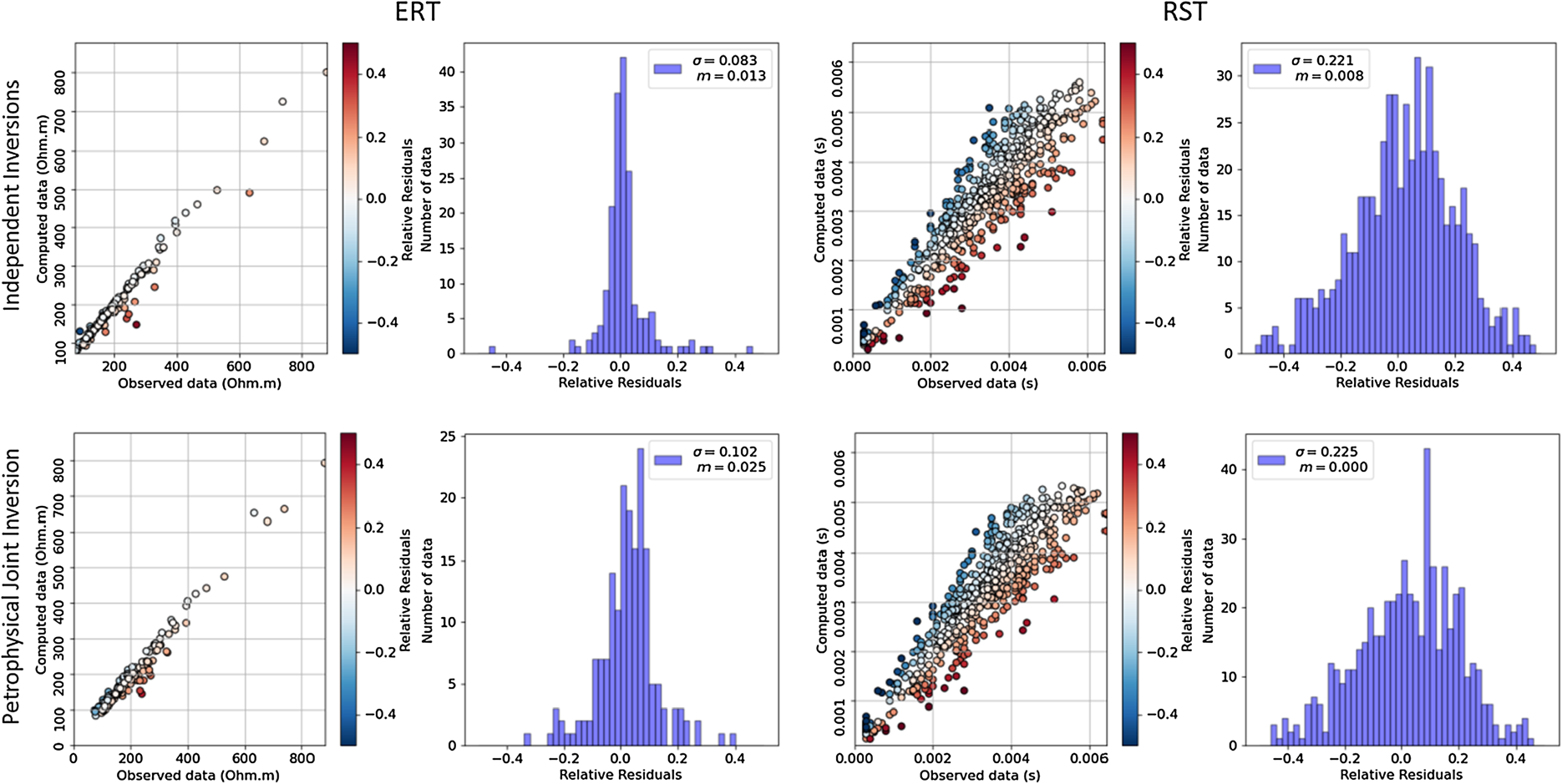}
\caption{CB06 pillar. Cross-plots (scatter plot, left) and residuals (histogram, right) are shown both for ERT (left column) and RST data (right column). First rows shows the Independent Inversion (IIv) and bottom row shows the Petrophysical Joint Inversion (P-JIv). The color scale of scatter points represents the relative residuals.}
\label{fig:qc06}
\end{figure}

\begin{figure}[htbp]
\centering
\includegraphics[width=\textwidth]{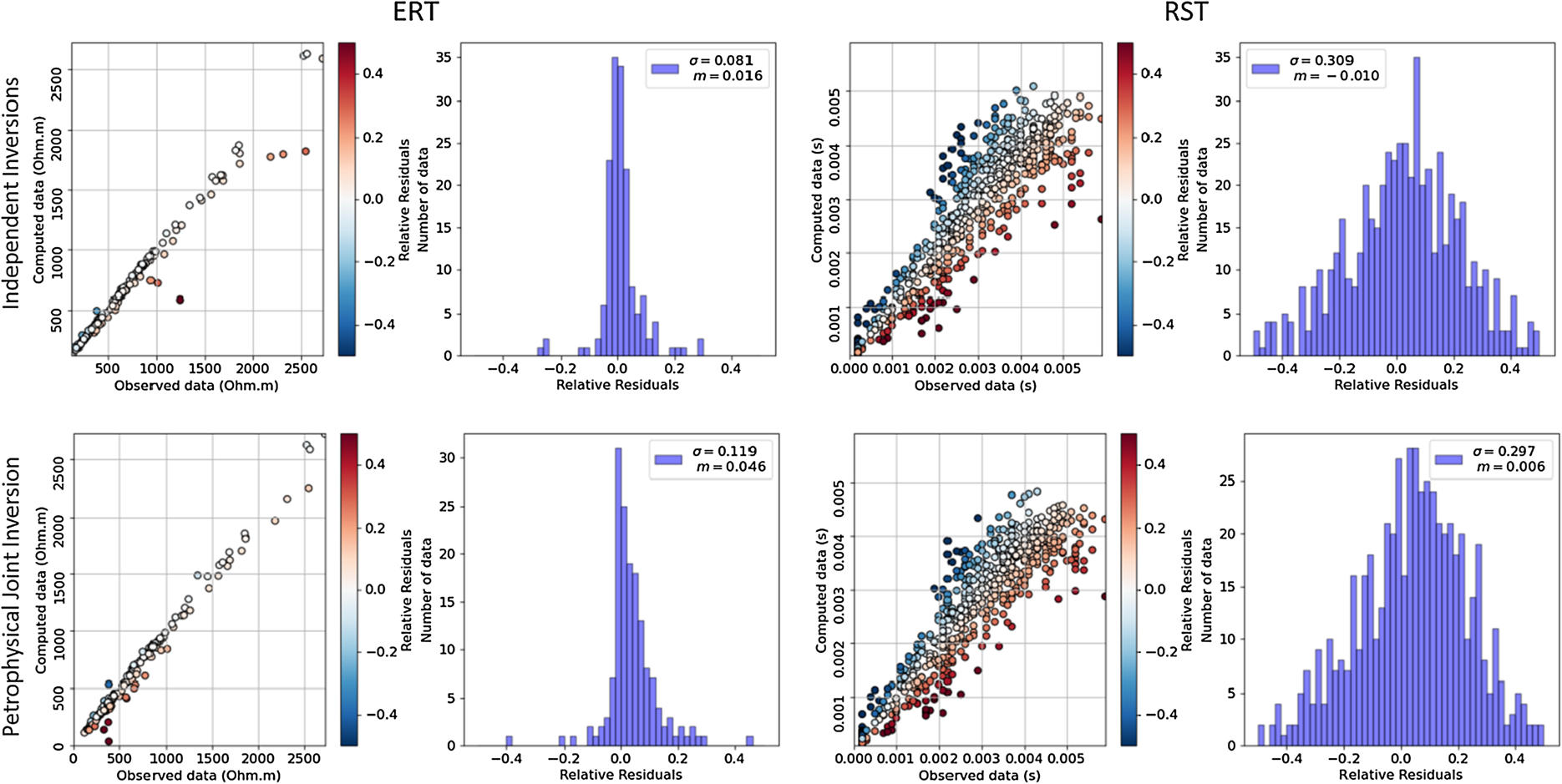}
\caption{CB10 pillar. Cross-plots (scatter plot, left) and residuals (histogram, right) are shown both for ERT (left column) and RST data (right column). First rows shows the Independent Inversion (IIv) and bottom row shows the Petrophysical Joint Inversion (P-JIv). The color scale of scatter points represents the relative residuals.}
\label{fig:qc10}
\end{figure}

Figures~\ref{fig:qc04}, \ref{fig:qc06} and \ref{fig:qc10} show the inversion quality indicators for CB04, CB06 and CB10, respectively. The top row shows the independent inversion results (IIv), bottom row shows the P-JIv inversion results. The left column shows the ERT indicators while RST statistics are shown in the right column. For each geophysical method and inversion scheme, the left panel shows the cross-plot between observed and computed data, while the right panel shows the residual distribution as a histogram.

Whatever the inversion scheme and method, the comparison between observed and computed data on the cross-plots is satisfactory. The distributions of the residuals are also properly centered, with broader distribution for RST than ERT (Figs.~\ref{fig:qc04}, \ref{fig:qc06}, \ref{fig:qc10}). Petrophysical joint inversion P-JIv tends to reduce the outliers in the residuals cross-plots. The mean and standard deviation of the residuals yet do not change much. The Relative RMS (RRMS) and $\chi^2$ show good inversion convergence, with slightly better results for ERT than RST (Table~\ref{tab:rms}). RST measurements indeed show greater relative errors because of the small scale of the investigated structures compared to picking/trigger accuracy. We also allowed the same low number of iterations (5 steps) for ERT and RST to ensure consistency between the independent and joint inversions. This might have stopped the RST inversion before full convergence. Whatever the studied pillar, joining ERT and RST inversions with P-JIv generally leads to a decrease in the RRMS and an increase in $\chi^2$ compared to IIv (see Table~\ref{tab:rms}). These values still lay in acceptable ranges. Note that the inversion results of individual and joint inversion are similar which is what is expected. Indeed, very different results would mean that either the ERT or RST methods are not valid taken individually or that the parameters for the P-JIv were not adapted.

\begin{table}[htbp]
\centering
\caption{Comparison of RST and ERT RRMS (\%) and $\chi^2$ (no unit) depending on the inversion scheme}
\label{tab:rms}
\begin{tabular}{llcccccc}
\toprule
 & & \multicolumn{3}{c}{RRMS (\%)} & \multicolumn{3}{c}{$\chi^2$} \\
\cmidrule(lr){3-5}\cmidrule(l){6-8}
 & & CB10 & CB06 & CB04 & CB10 & CB06 & CB04 \\
\midrule
\multirow{2}{*}{IIv} & ERT & 8.2 & 8.4 & 13.3 & 0.9 & 0.9 & 1.02 \\
 & RST & 29.9 & 21 & 38 & 6 & 6 & 6.7 \\
\multirow{2}{*}{P-JIv} & ERT & 5.3 & 8.8 & 4.8 & 3.1 & 8 & 2 \\
 & RST & 29.9 & 22.4 & 41 & 10.1 & 9 & 11 \\
\bottomrule
\end{tabular}
\end{table}

\begin{figure}[htbp]
\centering
\includegraphics[width=0.6\textwidth]{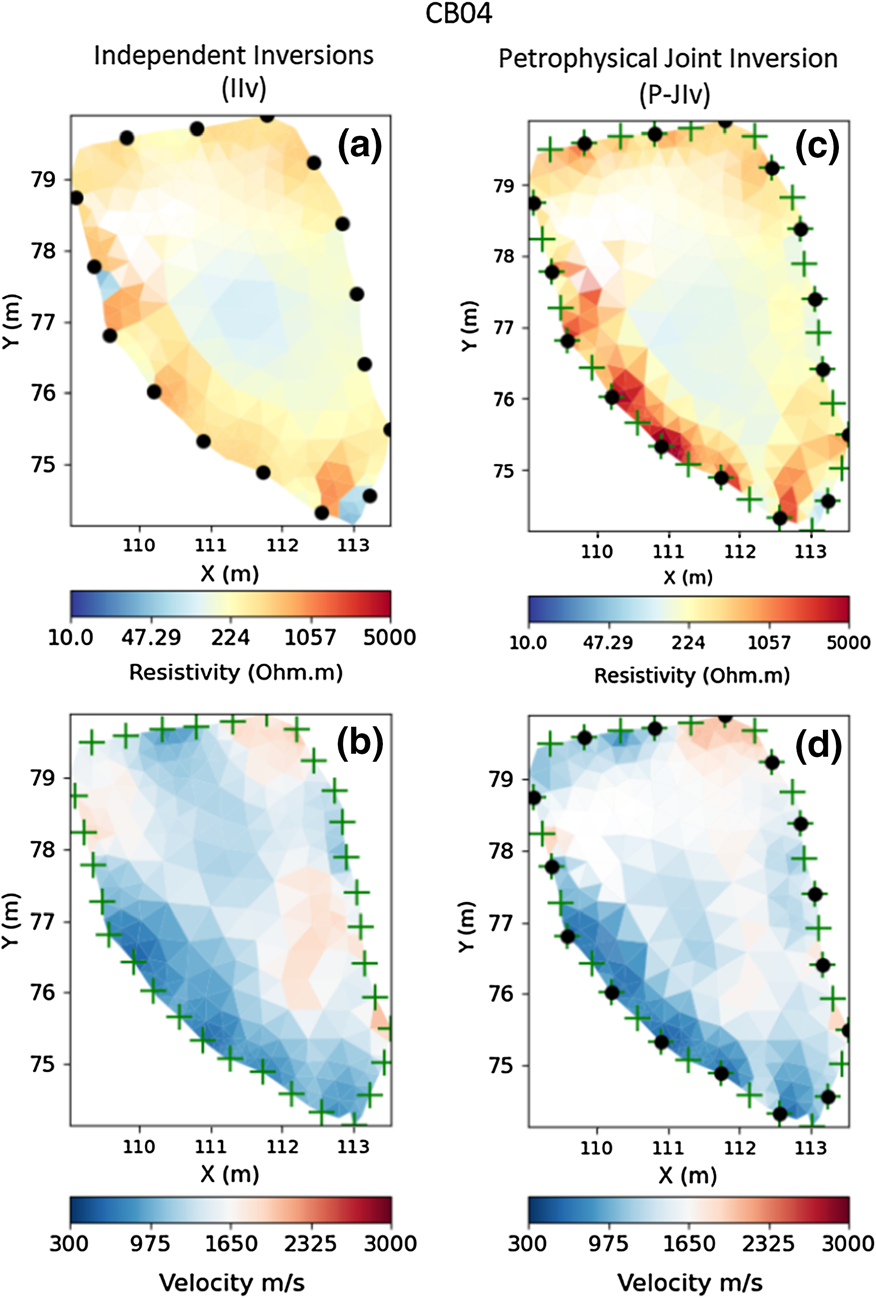}
\caption{Resistivity models (top row) and Vp velocity models (bottom) obtained after inversion on the CB04 pillar. Independent inversion (IIv) are shown on the left column (a and b insets) while Petrophysical Joint Inversion (P-JIv) is shown on the right column (c and d insets). Black circles: location of ERT sensors positions. Green crosses: RST sensors location.}
\label{fig:inv04}
\end{figure}

\begin{figure}[htbp]
\centering
\includegraphics[width=0.6\textwidth]{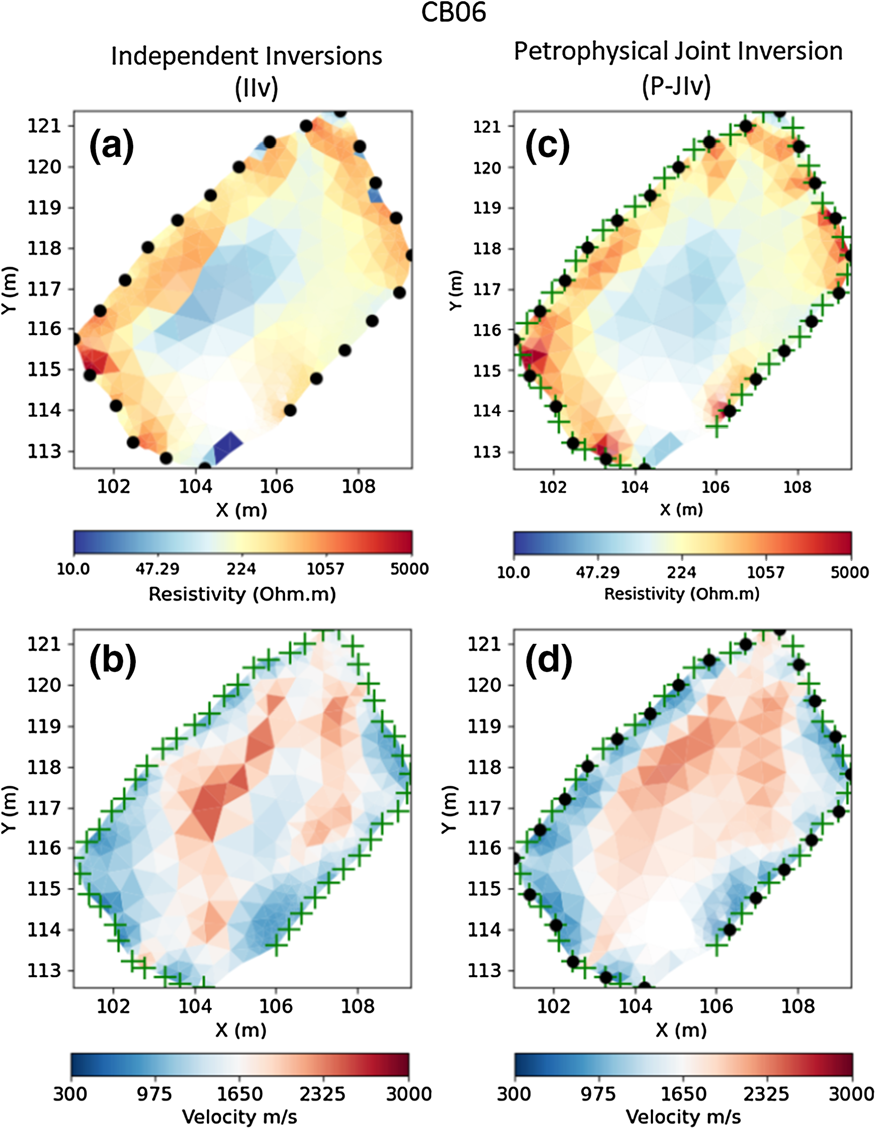}
\caption{Resistivity models (top row) and Vp velocity models (bottom) obtained after inversion on the CB06 pillar. Independent Inversion (IIv) are shown on the left column (a and b insets) while petrophysical joint inversion (P-JIv) is shown on the right column (c and d insets). Black circles: location of ERT sensors positions. Green crosses: RST sensors location.}
\label{fig:inv06}
\end{figure}

\begin{figure}[htbp]
\centering
\includegraphics[width=0.6\textwidth]{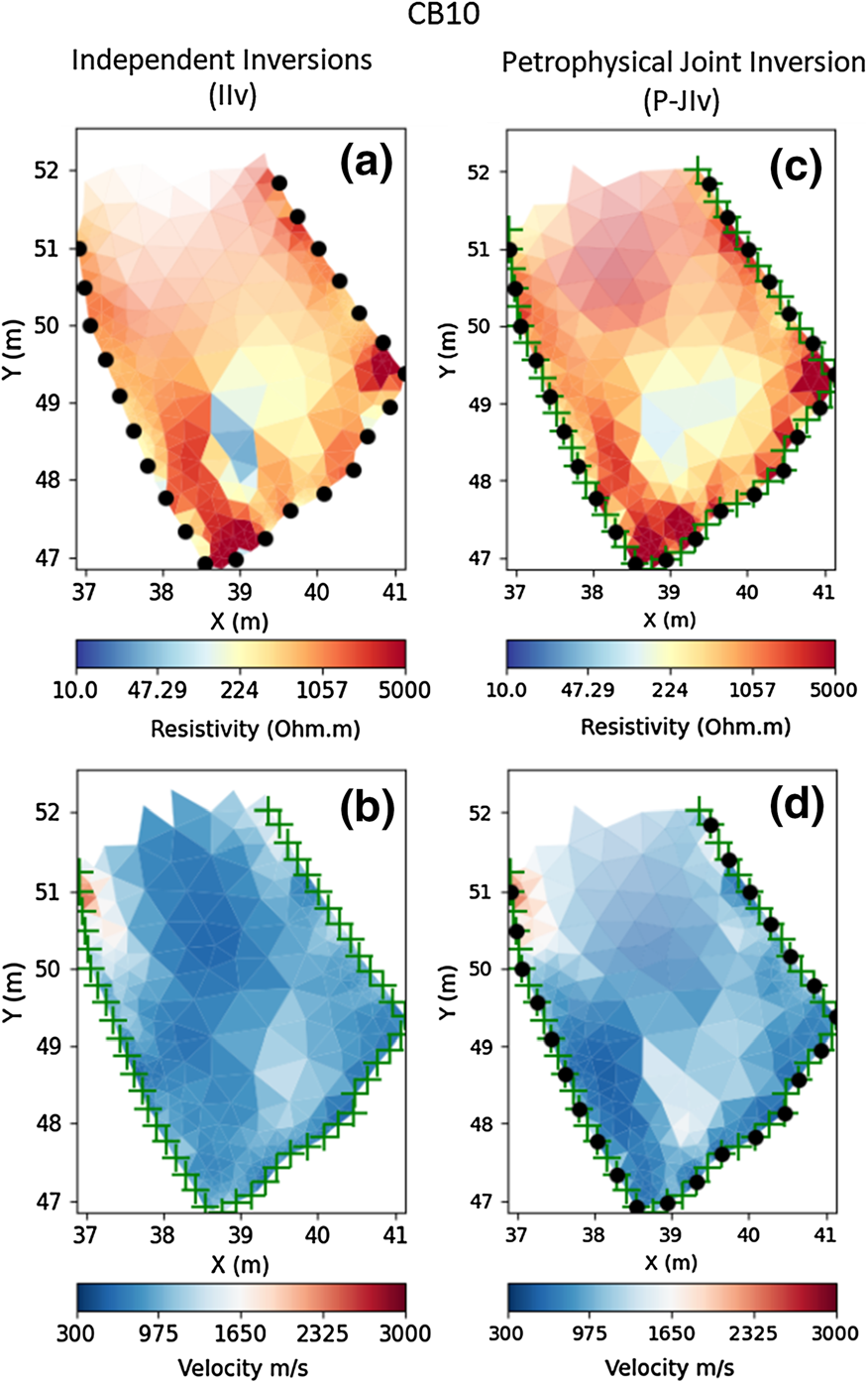}
\caption{Resistivity models (top row) and Vp velocity models (bottom) obtained after inversion on the CB10 pillar. Independent inversion (IIv) are shown on the left column (a and b insets) while Petrophysical Joint Inversion (P-JIv) is shown on the right column (c and d insets). Black circles: location of ERT sensors positions. Green crosses: RST sensors location.}
\label{fig:inv10}
\end{figure}

Electrical and seismic cross-sections for CB04, CB06 and CB10 pillars are shown in Figs.~\ref{fig:inv04}, \ref{fig:inv06} and \ref{fig:inv10}, respectively. Results from independent inversion (IIv) are shown in the left column and P-JIv results in the right. The same colorscales are used to map resistivity and seismic velocity, respectively, over all three pillars. Model sections are shaded according to their coverage.

For each pillar, ERT and RST sections obtained by IIv (left column) or P-JIv (right column) are very similar (Figs.~\ref{fig:inv04}, \ref{fig:inv06}, \ref{fig:inv10}). The major features of the models are very similar, which suggests a satisfactory inversion process and the presence of strong information into the data. P-JIv provides a slightly better spatial coverage in the results because each technique benefits from the coverage of the other. Whatever the pillar, the geophysical measurements appear very consistent with visual observations (see Sect.~2). Detailed result description is given below.

In the case of fully saturated chalky limestones, we expect fractured/damaged areas to be only partially saturated as fracturation would help evacuating fluids. The resistivity would be here increasing with damage due to the presence of air. Theoretically, damage areas should exhibits high resistivities and lower P-wave velocities. However, if the damaged rock is fully saturated resistivity will increase with a smaller effect on P-wave velocity, we expect here that joint inversion would help for interpretation through the estimation of rock fraction.

CB04 pillar shows the lowest overall resistivities among the three pillars. Outer first meter ring is resistive ($>1500$~Ohm~m) especially the south western face. The central core shows relatively conductive measurements ($<200$~Ohm~m). $V_p$ tomography is in very good agreement, with lower velocities on the western side ($<1000$~m/s) than the rest of the section (1500--2300~m/s). Indeed, on CB04, 20--40~cm fragments are observed on the south and western edges and the visual examination concluded to an overall healthy pillar.

CB06 exhibits quite conductive section ($<200$~Ohm~m) in its central part. The external 1~m thick fringe exhibits resistivities more than 200~Ohm~m and up to 5000~Ohm~m in NW corner. $V_p$ are relatively high, with the central core faster than 1500~m/s and up to more than 2000~m/s. The external 1~m thick ring exhibits velocities lower or equal to 1000~m/s. Such values suggest a good internal state of the pillar, with saturated (conductive) and compact rock. The visual inspection of the CB06 pillar suggests a solid and weakly damaged pillar, the main fragments are observed on the North-West and South-West edges of the pillars which is consistent with inversions results.

CB10 shows the highest resistivities (200--5000~Ohm~m) on almost its whole cross-section, except the lower central zone ($<100$~Ohm~m). The external border first meter in depth appears very resistive ($>4000$~Ohm~m). P-wave velocity is low ($<1500$~m/s on the whole cross-section, most of the time $<1000$~m/s). Such measurements suggest a very fractured, porous medium with low mechanical strength. The corners and edges appear strongly damaged. These results confirm the visual observations. Indeed, the latter highlights strongly fractured Northern, Eastern and Southern corners as well as geological fractures oriented N0, N120 and N140.

\begin{figure}[htbp]
\centering
\includegraphics[width=0.8\textwidth]{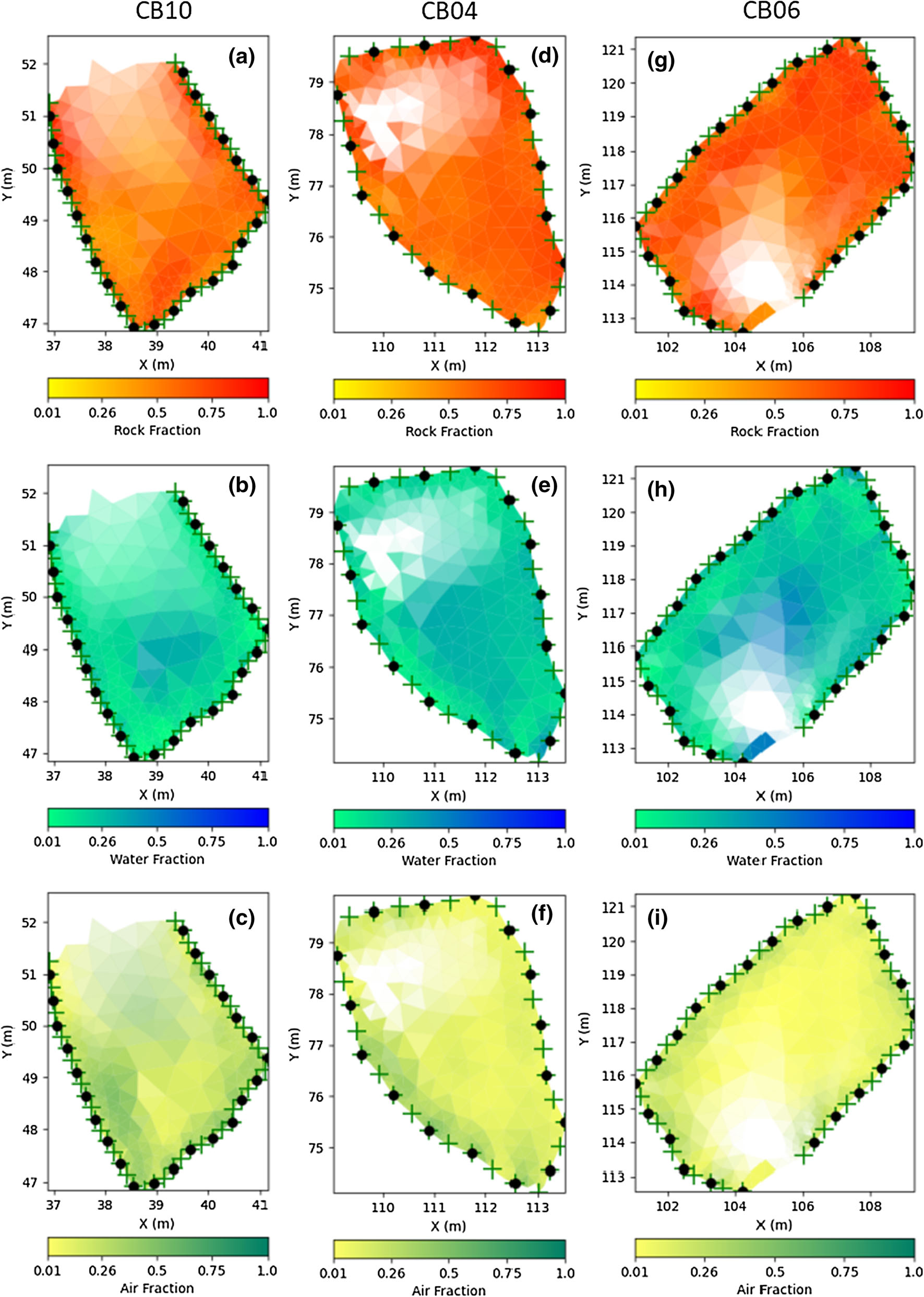}
\caption{Rock (top row), water (central row) and air fraction models (bottom row) obtained from Petrophysical Joint Inversion (P-JIv) on the CB04 (a--c), CB06 (d--f) and CB10 (g--i) pillars. The maps are shaded by the product of RST ray coverage and ERT coverage. Black circles: location of ERT sensors positions. Green crosses: RST sensors location.}
\label{fig:fractions}
\end{figure}

The main benefit provided by the use of a petrophysical model (see Sect.~\ref{sec:pjiv}) is the access provided to rock, water and air fractions (Eq.~\ref{eq:joint}). The rock, water and air fraction resulting from the P-JIv are shown in Fig.~\ref{fig:fractions} and shaded according to the coverage. As deducted from ERT and RST sections, CB10 shows the higher water and air fraction. This suggests a large, open porosity partially filled with air (i.e. fracturing). In contrast, CB04 and CB06 show negligible air fraction at the core. The rock porosity appears mostly filled with water, except at the western side of CB04 and northern, easter and southern sides of CB10 where air fraction is increasing.

\begin{figure}[htbp]
\centering
\includegraphics[width=\textwidth]{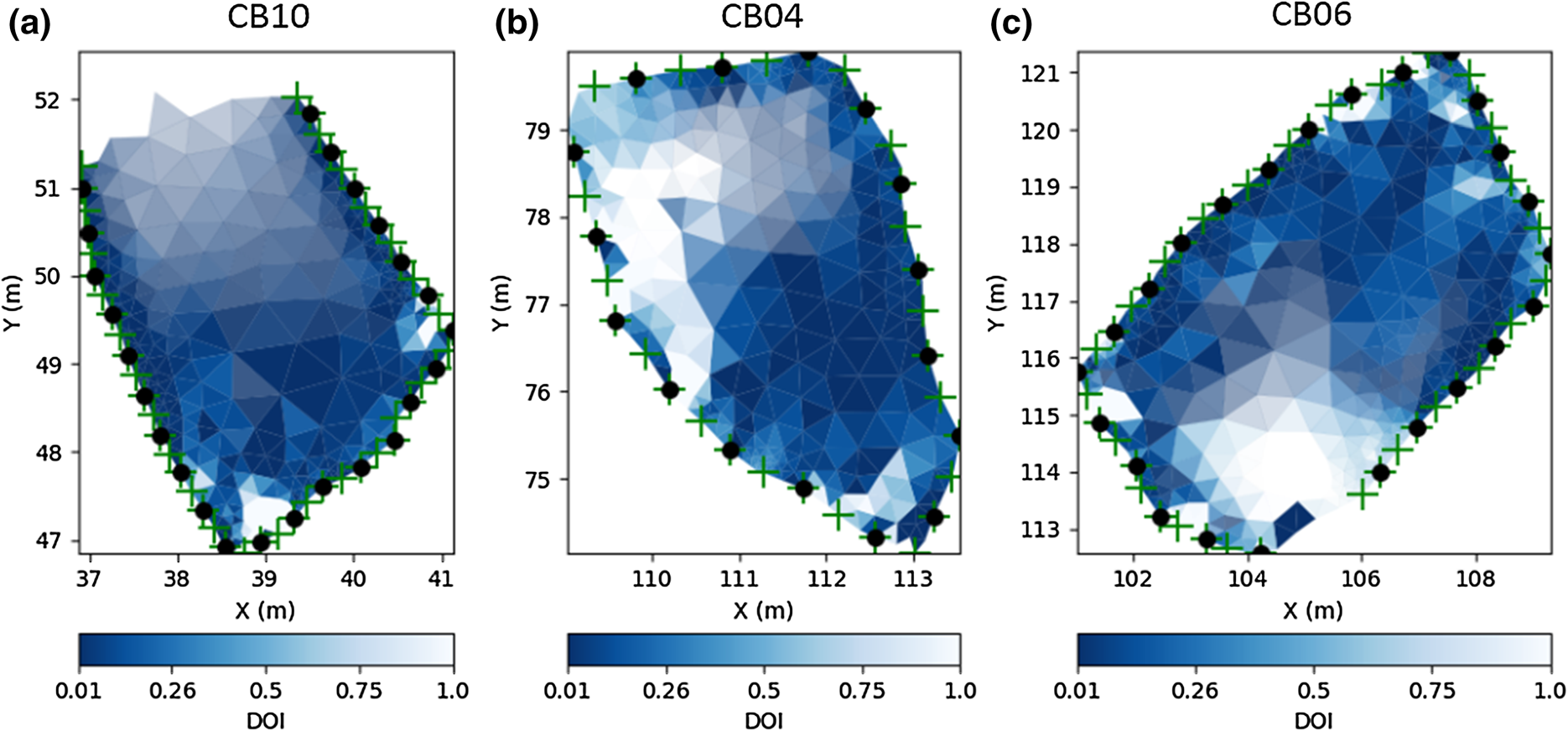}
\caption{Depth Of Investigation (DOI) index obtained from IIv inversions for CB10, CB04 and CB06 pillars, respectively. Black circles: location of ERT electrodes.}
\label{fig:doi}
\end{figure}

The DOI computed for individual ERT inversions is presented on Fig.~\ref{fig:doi}. It appears that areas where high and low resistivities are observed next to the other on CB04 and CB06 on the individual inversions are consistent with high DOI areas. It seems that here, data did not provide sufficient information to properly constrain the final model. These effects are smoothed in the joint inversion which suggests that joint inversion enables to compensate these artefacts.

The output models obtained via P-JIv are constrained by petrophysical relationships. These added constraints enable to decrease the number of outliers in the models and hence in the distribution of the residuals (c.f. Figs.~\ref{fig:qc04}--\ref{fig:inv10}).

\subsection{Comparison with Drilling Logs, Laboratory Tests and Geophysical Results}

The porosity obtained from P-JIv ranges between 20 and 60\%. For all the pillars the mean porosity equals or almost equals the median which means the distribution is close to be symmetrical. For CB04 and CB06, the porosity obtained from the inversion is $36 \pm 8$\% which is consistent with the laboratory results. For CB10, the porosity is higher and more widely distributed, i.e. $41 \pm 10$\%. The water content ranges from 1 to 52\% (and 3--98\% in the pores). The water content is $14 \pm 6$\% in CB10, $22 \pm 9$\% in CB06 and $20 \pm 6$\% in CB04, higher values being observed at the core of the pillars ($35 \pm 14$\% in CB10, $60 \pm 20$\% in CB06 and $55 \pm 20$\% in CB04 in the pores). This suggests that voids are more easily drained in CB10 than in the other pillars.

\begin{figure}[htbp]
\centering
\includegraphics[width=\textwidth]{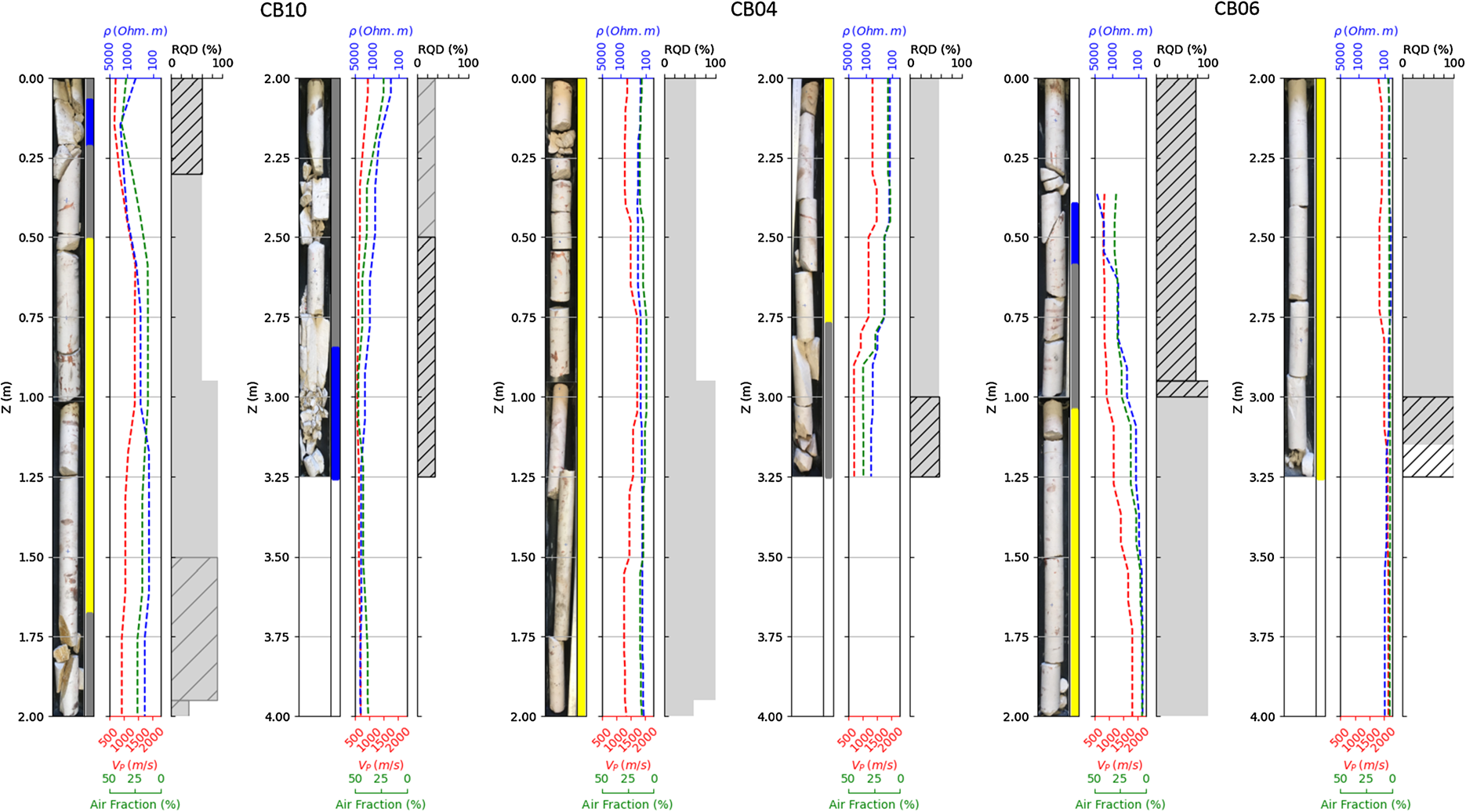}
\caption{Comparison between drilling cores (left pictures), rock quality designation RQD (right bar graph) and geophysical model (central panel) for CB10, CB04 and CB06. The central graph shows the seismic velocity (red dashed line), electrical resistivity (blue dashed line) and air fraction (green dashed lines). The cluster groups extracted from P-JIv are shown in the right margin of drilling log pictures. Cluster groups 0, 1 and 2 are shown in yellow, grey and blue, respectively. For RQD, the gray bars corresponds to RQD distribution computed on 1~m cores samples. The black striped areas indicate fracture zones while gray striped areas show moderately fractured areas with alteration traces.}
\label{fig:logs}
\end{figure}

ERT ($\rho$) and RST ($V_P$) maps from P-JIv were interpolated along the borehole locations. They are shown as blue and red dashed lines, respectively in Fig.~\ref{fig:logs}. Note that the resistivity axis is reversed. Visual observations are added through dashed coded rectangles. Calcite veins are observed in good quality rock cores. Oxidation deposits can be observed in moderately damaged areas (grey shaded areas). We observe a very good consistency between geophysical results and drilling cores. Areas with good rock quality (high RQD, massive appearance) clearly correspond to peaks in $V_P$ and troughs in $\rho$. Inversely, fractured areas (hatched areas with low RQD) show clear decrease in $V_P$ and increase in $\rho$. The air fraction (green dashed line)---proxy of rock fracturing (c.f. Sect.~\ref{sec:pjiv})---also shows a very good agreement with drilling results. The location of the high air content areas is consistent with visual inspection and fracture observations (black dashed areas on Fig.~\ref{fig:logs}).

\subsection{Clustering Results}

\begin{figure}[htbp]
\centering
\includegraphics[width=0.5\textwidth]{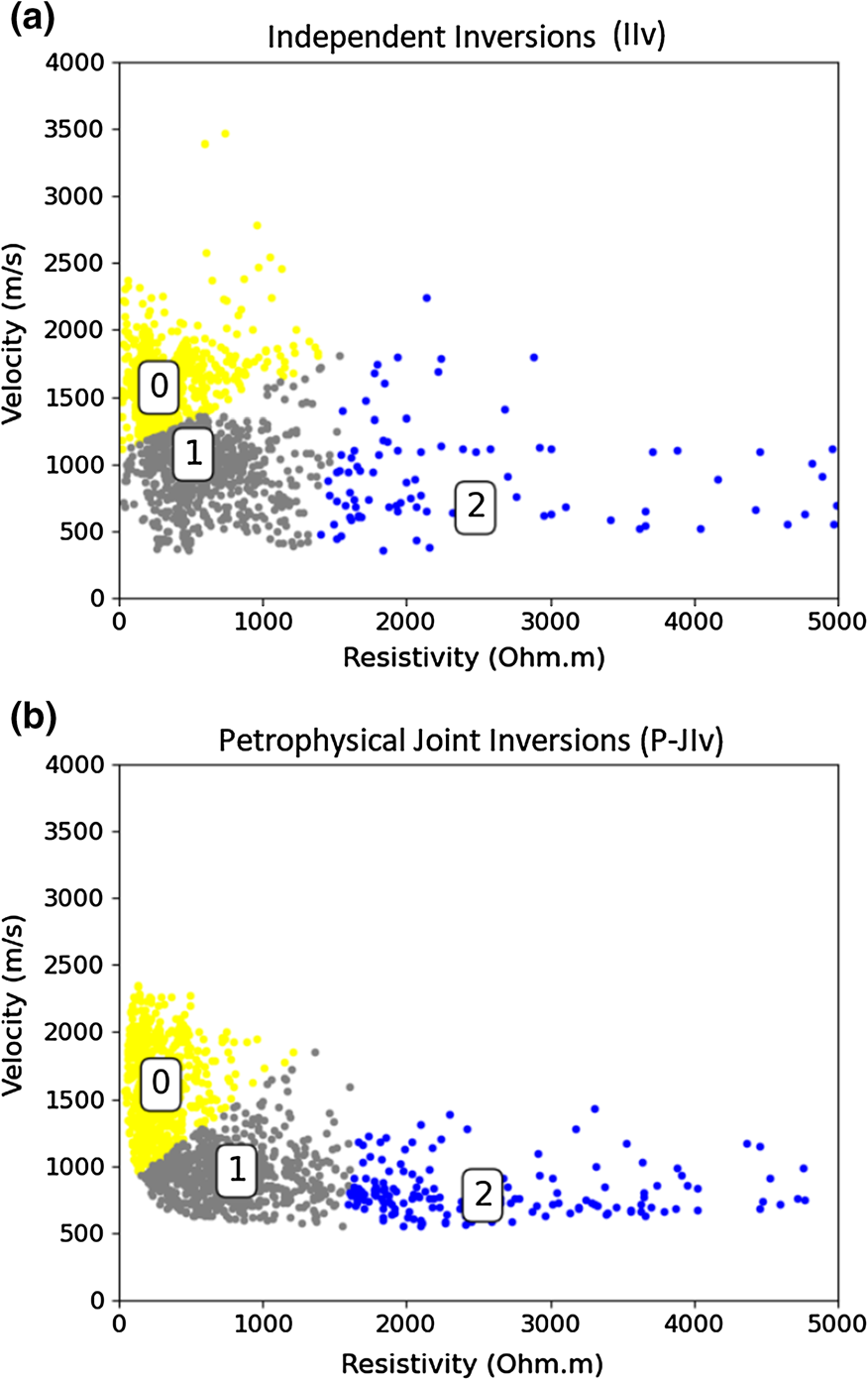}
\caption{Clustering results for (resistivity, seismic velocity) data. a Independent inversions (IIv) and b petrophysical joint inversions (P-JIv). The number of clusters has been forced to 3, labeled cluster 0 (yellow), cluster 1 (grey) and cluster 2 (blue).}
\label{fig:clusters}
\end{figure}

\begin{figure}[htbp]
\centering
\includegraphics[width=\textwidth]{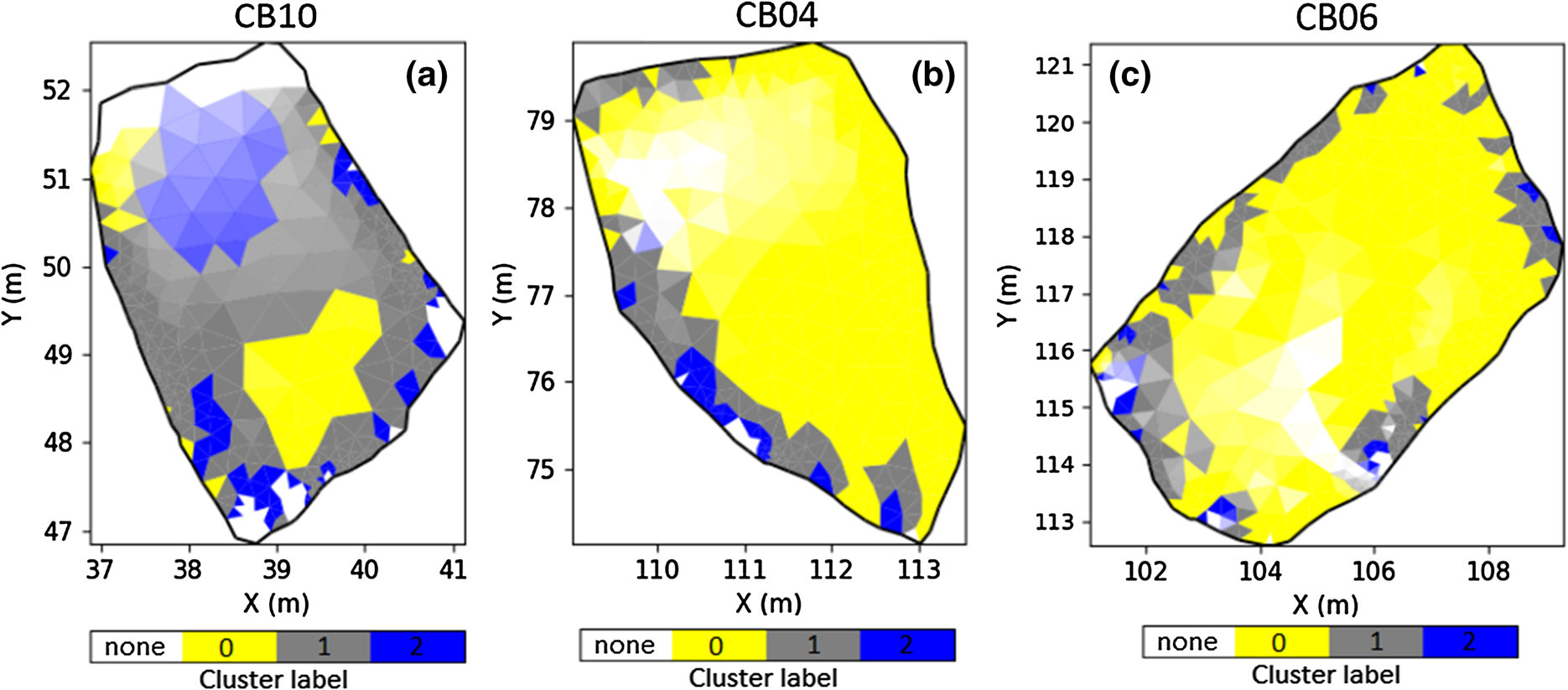}
\caption{Spatial distribution of the (resistivity, seismic velocity) clusters for the Petrophysical Joint Inversion (P-JIv). The maps are shaded by the product of RST ray coverage and ERT coverage. Cluster groups 0, 1 and 2 are shown in yellow, grey and blue, respectively. Unclassified areas are left blank.}
\label{fig:clustmap}
\end{figure}

The [resistivity-seismic velocity] values were grouped for all three pillars and clustered using the process described in Sect.~\ref{sec:clust}. The clustering has been conducted on both IIv and P-JIv results (Fig.~\ref{fig:clusters}, top and bottom respectively). The results from P-JIv shows more focused [resistivity-seismic velocity] distribution than IIv. Hence, we considered only the P-JIv clustering results hereafter. Three groups of data labeled 0, 1 and 2 were automatically drawn in Fig.~\ref{fig:clusters} along with group centroids. Based on resistivity and seismic velocity values observed for each group, we reasonably established that group 0 corresponds to rock in good condition, group 1 depicts intermediate state rock and group 2 represents very weak rock. The spatial distribution of the clusters is presented on Fig.~\ref{fig:clustmap}. Pillars CB04 and CB06 reveal degraded material on the edges and rims only and exhibit good rock quality cores. On the contrary, CB10 shows mostly intermediate and very weak rock.

Although such classification seems arbitrary and relevant only on a given site for a given material, it allows deriving objective comparisons between pillars. As an example, Table~\ref{tab:clust} shows the surface percentage of each cluster for each pillar. CB04 and CB06 appear to have more than 78\% of healthy material as opposed to CB10 which exhibits only 16\% of good state rock.

Cluster identity numbers are also interpolated along the borehole locations. Good quality rocks---higher velocity and lower resistivity values---are associated to cluster number 0 (yellow). No fractures observation are made in cores located in cluster 0 area. Observed damaged areas (black dashed areas) are associated to clusters 1 and 2 (gray and blue). All the results are consistent with visual observation and drilling logs and enable to better quantify the damage degree of the pillar and to evaluate the extent of the damage observed on the surface of the pillars. CB10 appears to be the most damaged pillar with damage areas reaching the center of the pillar. CB06 and CB04 consist in healthier pillars with a undamaged core and damage limited to the edges and corners.

\begin{table}[htbp]
\centering
\caption{Percentage of healthy, intermediate and altered rock surface on each pillar extracted from clustering results on P-JIv}
\label{tab:clust}
\begin{tabular}{lccc}
\toprule
 & CB04 & CB06 & CB10 \\
\midrule
Healthy rock (cluster 0) & 79.3 & 78.2 & 16.2 \\
Intermediate (cluster 1) & 16.3 & 18.2 & 51.1 \\
Degraded (cluster 2) & 3.8 & 2.6 & 22.7 \\
No cover & 0.6 & 1 & 10 \\
\bottomrule
\end{tabular}

\smallskip
\parbox{0.6\textwidth}{\footnotesize Areas with no resolution are labeled as no cover.}
\end{table}

\section{Discussion}

The resistivity versus velocity scatterplots for joint inversion show less scattered distribution than individual inversions (Fig.~\ref{fig:clusters}). This suggests that the use of joint inversions yielded more focused results in terms of resistivity and velocity models. As a result, the zonation resulting from the clustering is clearer and the identification of the degree of damage is more reliable. However, petrophysical joint inversion P-JIv needs a good understanding of the local geology and some a priori knowledge on the petrophysical properties of the considered rocks. Petrophysical joint inversion indeed implies the use of petrophysical relationships and its applicability range lies in the applicability range of Archie's and Wyllie's laws. What's more, the use of parameters such as rock velocity, fluid resistivity, saturation and cementation exponents imply to have some laboratory data or a good geological and petrophysical knowledge of the area. In our case, we had to approximate these values, which implies that the results of the models may not be truly quantitative.

More recently, Mollaret et al. (2020) compared the used of different petrophysical relationships, implying or not the addition of surface conductivity, for joint inversion models for permafrost applications. The resulting models obtained with Archie's law and modified Archie's law and purely surface conduction lead to similar results. They explained those similarities by the lack of laboratory data to constrained the parameters of the petrophysical models and the consequent degree of liberty of the free parameters. What's more, they show that the use of different $m$ or $n$ parameters lead to similar RMS and $\chi^2$ values making difficult to choose petrophysical parameters only by the reliability of the inversion. They also observed that the final reliability of the models (with or without surface conductivity included) depends on the initial porosity estimate which should be close to reality.

Considering the interpretation of the air fraction, we choose a very simplified point of view which does not take into account petrophysical specifities nor chemical reactions that could occur at the surface of the pillars while drying. For example, we did not consider the fact that at the surface of the pillars, drying might imply pore water conductivity variations and possibly a strong heterogeneity. In our case, it is difficult to discriminate between drying and water draining through fractures, considering ambiant air humidity, we firstly choose to neglect this fact.

For RST inversions, we noticed that misfits are relatively high. This results probably from an underestimation of the data error in the inversion due to perfectible triggering time accuracy. These uncertainties make difficult a precise interpretation of the velocity sections obtained in a quantitative way. Observed velocity contrasts and zonation are yet correct and realistic. Due to these difficulties, we choose to set a low maximum number of iterations (5 iterations) in order to avoid overfitting of the data. This may lead to smoother but robust velocity models.

Spatial zonation of the parameters through clustering computation enables to better identify the different areas according to their damage degree without the need to quantify the parameters of the models. The simplified results are easier to read and to interpret and damage degree quantification is hence clearer and make a more readable release for non-specialists (Fig.~\ref{fig:clustmap}). We believe that such automatic classification algorithm may represent a substantial help indicator for decision, especially when a large number of geophysical sections have to be interpreted together.

\section{Conclusion}

The use of refraction seismic and electrical resistivity tomography has been tested in order to evaluate the damage degree of supporting pillars in an underground cave. Petrophysical joint inversion method (P-JIv) has been tested and compared to the results of individual inversions (IIv).

The results suggest that the use of combined RST and ERT enables to identify damage areas in the pillars, whatever the inversion technique. The implementation of joint inversion (JIv) yields more consistent results and less scattered distribution of the resulting velocity and resistivity models than independent inversions (IIv). Geophysical results are very consistent with visual inspection of the subsurface of the pillars, with drillings logs results and with cores observations. Petrophysical joint inversion (P-JIv) enabled to get access to the fraction of air, rock and water in the pillars. It also enables to decrease inversion artefacts in areas where data from one of the two datasets do not provide information in the individual inversion processes. Hence, the overall coverage and consistency is improved as each method benefits from the coverage of the other. However, the air, rock and water content obtained are closely linked to the petrophysical laws chosen to perform the joint inversion and the approximations made should be always kept in mind. To better estimate the importance of fracturing, a better knowledge of the porosity distribution and characteristics in intact rocks would be needed. Also, the understanding of the drying processes at the border of the pillars could be useful.

The spatial zonation of the parameters through a clustering approach yielded clear and easy-to-read results even for non specialists. This approach allowed to map several areas according to their damage degree and give a quantitative damage indicator helping in prioritizing the reinforcement works. We believe that such clustering techniques may represent a powerful help for decision indicator but should always be preceded by a careful examination of datasets and results.

\section*{Acknowledgements}

The authors express their gratitude to Benoît Michon who carried out the drilling operations in the cave, and Sylvie Ustal, Tracy Decotte and Vincent Faudou who performed the laboratory tests on rock samples. Thanks to the winery for facilitated access to the premises and the aerial bucket provided, especially to responsible for his help and arrangement. We also would like to thank Florian Wagner for providing the 4PM codes, as well as Thomas Günther and Carsten Rücker for providing and maintaining the pygimli python package (all the source codes are available via the pygimli website or GitHub). The authors also warmfully thank the two anonymous reviewers for their helpful comments that greatly improved the quality of the manuscript.

\section*{Declarations}

\textbf{Consent for publication.} The authors have the consent for publication of the data from the studied quarry under the condition of anonymity.

\section*{References}
\begingroup
\small
\setstretch{1.1}
\setlength{\parindent}{-1.5em}
\setlength{\leftskip}{1.5em}
\setlength{\parskip}{0.3em}

Al Heib, M., Duval, C., Theoleyre, F., Watelet, J.-M., \& Gombert, P. (2015). Analysis of the historical collapse of an abandoned underground chalk mine in 1961 in clamart (Paris, France). \emph{Bulletin of Engineering Geology and the Environment}, 74(3), 1001--1018.

Archie, G. E., et al. (1942). The electrical resistivity log as an aid in determining some reservoir characteristics. \emph{Transactions of the AIME}, 146(01), 54--62.

Bakun-Mazor, D., Hatzor, Y. H., \& Dershowitz, W. S. (2009). Modeling mechanical layering effects on stability of underground openings in jointed sedimentary rocks. \emph{International Journal of Rock Mechanics and Mining Sciences}, 46(2), 262--271.

Bartolo, S. D., \& Salvini, R. (2019). Multitemporal terrestrial laser scanning for marble extraction assessment in an underground quarry of the apuan alps (italy). \emph{Sensors}, 19(3), 450. \url{https://doi.org/10.3390/s19030450}.

Bedrosian, P., Maercklin, N., Weckmann, U., Bartov, Y., Ryberg, T., \& Ritter, O. (2007). Lithology-derived structure classification from the joint interpretation of magnetotelluric and seismic models. \emph{Geophysical Journal International}, 170(2), 737--748.

Bolève, A., Crespy, A., Revil, A., Janod, F., \& Mattiuzzo, J.-L. (2007). Streaming potentials of granular media: Influence of the Dukhin and Reynolds numbers. \emph{Journal of Geophysical Research: Solid Earth}, 112(B8), 1--14.

Brady, B. H. \& Brown., E. T. (2006). \emph{Rock Mechanics for underground mining}. Springer.

BRGM. (2015). Fiche de caracterisation MEHG307, cycle DCE 2016--2021, calcaires Kimmeridgien-Oxfordien karstique entre Yonne et Seine. BRGM: Tech. rep.

Broere, W. (2016). Urban underground space: Solving the problems of today's cities. \emph{Tunnelling and Underground Space Technology} 55, 245--248, urban Underground Space: A Growing Imperative Perspectives and Current Research in Planning and Design for Underground Space Use.

Brownjohn, J. M. (2007). Structural health monitoring of civil infrastructure. \emph{Philosophical Transactions of the Royal Society A: Mathematical, Physical and Engineering Sciences}, 365(1851), 589--622.

Brunet, P., Clément, R., \& Bouvier, C. (2010). Monitoring soil water content and deficit using electrical resistivity tomography (ert)-a case study in the cevennes area, france. \emph{Journal of Hydrology}, 380(1--2), 146--153.

Cała, M., Stopkowicz, A., Kowalski, M., Blajer, M., Cyran, K., \& D'obyrn, K. (2016). Stability analysis of underground mining openings with complex geometry. \emph{Studia Geotechnica et Mechanica}, 38(1), 25--32.

Cardarelli, E., Cercato, M., Cerreto, A., \& Di Filippo, G. (2010). Electrical resistivity and seismic refraction tomography to detect buried cavities. \emph{Geophysical Prospecting}, 58(4), 685--695.

Carrier, A., Fischanger, F., Gance, J., Cocchiararo, G., Morelli, G., \& Lupi, M. (2019). Deep electrical resistivity tomography for the prospection of low-to medium-enthalpy geothermal resources. \emph{Geophysical Journal International}, 219(3), 2056--2072.

Caterina, D., Beaujean, J., Robert, T., \& Nguyen, F. (2013). A comparison study of different image appraisal tools for electrical resistivity tomography. \emph{Near Surface Geophysics}, 11(6), 639--658.

Comaniciu, D., \& Meer, P. (2002). Mean shift: A robust approach toward feature space analysis. \emph{IEEE Transactions on Pattern Analysis and Machine Intelligence}, 24(5), 603--619.

Coperey, A., Revil, A., \& Stutz, B. (2019). Electrical conductivity versus temperature in freezing conditions: A field experiment using a basket geothermal heat exchanger. \emph{Geophysical Research Letters}, 46, 14531--14538.

Cote, P., Dérobert, X., \& Abraham, O. (2005). Typology of conditions for underground quarry pillars based on geophysical investigations. In: \emph{Int. Symp. Post-Mining}.

Coutant, O., Bernard, M., Beauducel, F., Nicollin, F., Bouin, M., \& Roussel, S. (2012). Joint inversion of P-wave velocity and density, application to La Soufrière of Guadeloupe hydrothermal system. \emph{Geophysical Journal International}, 191(2), 723--742.

Das, P., Pal, S., Mohanty, P., Priyam, P., Bharti, A. K., \& Kumar, R. (2017). Abandoned mine galleries detection using electrical resistivity tomography method over Jharia coal field, India. \emph{Journal of the Geological Society of India}, 90(2), 169--174.

Dérobert, X., \& Abraham, O. (2000). Gpr and seismic imaging in a gypsum quarry. \emph{Journal of Applied Geophysics}, 45(3), 157--169.

Dessandier, D., Ambroise, G., \& Longet, A. (2006). Mémento des pierres ornementales et de construction de la région bourgogne (p. BRGM/RP-54618-FR.). BRGM: Tech. rep.

Díaz, D., Maksymowicz, A., Vargas, G., Vera, E., Contreras-Reyes, E., \& Rebolledo, S. (2014). Exploring the shallow structure of the san ramón thrust fault in Santiago, Chile (33.5 s), using active seismic and electric methods. \emph{Solid Earth}, 5(2), 837.

Doetsch, J., Linde, N., Coscia, I., Greenhalgh, S. A., \& Green, A. G. (2010). Zonation for 3d aquifer characterization based on joint inversions of multimethod crosshole geophysical data. \emph{Geophysics}, 75(6), G53--G64.

Doetsch, J., Linde, N., Pessognelli, M., Green, A. G., \& Günther, T. (2012). Constraining 3-d electrical resistance tomography with GPR reflection data for improved aquifer characterization. \emph{Journal of Applied Geophysics}, 78, 68--76.

Draebing, D., \& Krautblatter, M. (2012). P-wave velocity changes in freezing hard low-porosity rocks: A laboratory-based time-average model. \emph{The Cryosphere}, 6(5), 1163--1174.

Duvillard, P., Revil, A., Qi, Y., Soueid Ahmed, A., Coperey, A., \& Ravanel, L. (2018). Three-dimensional electrical conductivity and induced polarization tomography of a rock glacier. \emph{Journal of Geophysical Research: Solid Earth}, 123(11), 9528--9554.

Eberhart-Phillips, D., Stanley, W. D., Rodriguez, B. D., \& Lutter, W. J. (1995). Surface seismic and electrical methods to detect fluids related to faulting. \emph{Journal of Geophysical Research: Solid Earth}, 100(B7), 12919--12936.

Fabricius, I. L. (2020). Porosity in chalk-roles of elastic strain and plastic strain. \emph{Sedimentology}, 67(7), 3451--3470.

Fargier, Y., Antoine, R., Dore, L., Lopes, S. P., \& Fauchard, C. (2017). 3d assessment of an underground mine pillar by combination of photogrammetric and geoelectric methods: combination of photogrammetry and 3d-eri. \emph{Geophysics}, 82(4), E143--E153.

Gallardo, L. A., \& Meju, M. A. (2003). Characterization of heterogeneous near-surface materials by joint 2d inversion of dc resistivity and seismic data. \emph{Geophysical Research Letters}, 30(13), 1--4.

Gallardo, L. A., \& Meju, M. A. (2004). Joint two-dimensional dc resistivity and seismic travel time inversion with cross-gradients constraints. \emph{Journal of Geophysical Research: Solid Earth}, 109(B3), 1--11.

Gallardo, L. A., \& Meju, M. A. (2011). Structure-coupled multiphysics imaging in geophysical sciences. \emph{Reviews of Geophysics}, 49(1), 1--19.

Ganerød, G. V., Rønning, J. S., Dalsegg, E., Elvebakk, H., Holmøy, K., Nilsen, B., \& Braathen, A. (2006). Comparison of geophysical methods for sub-surface mapping of faults and fracture zones in a section of the viggja road tunnel, norway. \emph{Bulletin of Engineering Geology and the Environment}, 65(3), 231.

Glover, P. W. (2010). A generalized Archie's law for n phases. \emph{Geophysics}, 75(6), E247--E265.

Gunther, T., \& Rucker, C. (2006). A new joint inversion approach applied to the combined tomography of dc resistivity and seismic refraction data. In: \emph{Symposium on the Application of Geophysics to Engineering and Environmental Problems}. Society of Exploration Geophysicists, pp. 1196--1202.

Günther, T., Rücker, C., \& Spitzer, K. (2006). Three-dimensional modelling and inversion of dc resistivity data incorporating topography-ii inversion. \emph{Geophysical Journal International}, 166(2), 506--517.

Hanesse, M., \& Richard, T. (2016). Guide de surveillance des cavités souterraines d'anthropique. Guide technique. Rapport INERIS (p. DRS-16-156834-00810B, 61.)

Hauck, C., Böttcher, M., \& Maurer, H. (2011). A new model for estimating subsurface ice content based on combined electrical and seismic data sets. \emph{The Cryosphere}, 5(2), 453--468.

Hellman, K., Ronczka, M., Günther, T., Wennermark, M., Rücker, C., \& Dahlin, T. (2017). Structurally coupled inversion of ERT and refraction seismic data combined with cluster-based model integration. \emph{Journal of Applied Geophysics}, 143, 169--181.

Hermans, T., Vandenbohede, A., Lebbe, L., \& Nguyen, F. (2012). A shallow geothermal experiment in a sandy aquifer monitored using electric resistivity tomography. \emph{Geophysics}, 77(1), B11--B21.

Hilbich, C., Marescot, L., Hauck, C., Loke, M., \& Mäusbacher, R. (2009). Applicability of electrical resistivity tomography monitoring to coarse blocky and ice-rich permafrost landforms. \emph{Permafrost and Periglacial Processes}, 20(3), 269--284.

Hirsch, M., Bentley, L. R., \& Dietrich, P. (2008). A comparison of electrical resistivity, ground penetrating radar and seismic refraction results at a river terrace site. \emph{Journal of Environmental \& Engineering Geophysics}, 13(4), 325--333.

Idrees, M. O., \& Pradhan, B. (2018). Geostructural stability assessment of cave using rock surface discontinuity extracted from terrestrial laser scanning point cloud. \emph{Journal of Rock Mechanics and Geotechnical Engineering}, 10(3), 534--544.

Jain, A. K., Duin, R. P. W., \& Mao, J. (2000). Statistical pattern recognition: A review. \emph{IEEE Transactions on Pattern Analysis and Machine Intelligence}, 22(1), 4--37.

Jones, G., Sentenac, P., \& Zielinski, M. (2014). Desiccation cracking detection using 2-d and 3-d electrical resistivity tomography: Validation on a flood embankment. \emph{Journal of Applied Geophysics}, 106, 196--211.

Kortnik, J. (2015). Stability assessment of the high safety pillars in Slovenian natural stone mines. \emph{Archives of Mining Sciences}, 60(1), 403--417.

Kumar, R., Pal, S. K., \& Gupta, P. K. (2021). Water seepage mapping in an underground coal-mine barrier using self-potential and electrical resistivity tomography. \emph{Mine Water and the Environment}, 1--17.

Lafrance, N., Auvray, C., Souley, M., \& Labiouse, V. (2016). Impact of weathering on macro-mechanical properties of chalk: Local pillar-scale study of two underground quarries in the paris basin. \emph{Engineering Geology}, 213, 107--119.

Lajaunie, M., Gance, J., Nevers, P., Malet, J., Bertrand, C., Garin, T., \& Ferhat, G. (2019). Structure of the séchilienne unstable slope from large-scale three-dimensional electrical tomography using a resistivity distributed automated system (r-das). \emph{Geophysical Journal International}, 219(1), 129--147.

Lelièvre, P. G., Farquharson, C. G., \& Hurich, C. A. (2012). Joint inversion of seismic traveltimes and gravity data on unstructured grids with application to mineral exploration. \emph{Geophysics}, 77(1), K1--K15.

Leroy, P., Revil, A., Kemna, A., Cosenza, P., \& Ghorbani, A. (2008). Complex conductivity of water-saturated packs of glass beads. \emph{Journal of Colloid and Interface Science}, 321(1), 103--117.

Lesparre, N., Grychtol, B., Gibert, D., Komorowski, J.-C., \& Adler, A. (2014). Cross-section electrical resistance tomography of la soufrière of guadeloupe lava dome. \emph{Geophysical Journal International}, 197(3), 1516--1526.

Leucci, G., Greco, F., De Giorgi, L., \& Mauceri, R. (2007). Three-dimensional image of seismic refraction tomography and electrical resistivity tomography survey in the castle of occhiola (sicily, italy). \emph{Journal of Archaeological Science}, 34(2), 233--242.

Li, J., Liu, C., Zeng, Z., \& Chen, L. (2015). Gpr signal denoising and target extraction with the CEEMD method. \emph{IEEE Geoscience and Remote Sensing Letters}, 12(8), 1615--1619.

Linde, N., Binley, A., Tryggvason, A., Pedersen, L. B., \& Revil, A. (2006). Improved hydrogeophysical characterization using joint inversion of cross-hole electrical resistance and ground-penetrating radar traveltime data. \emph{Water Resources Research}, 42(12), 1--16.

Linder, S., Paasche, H., Tronicke, J., Niederleithinger, E., \& Vienken, T. (2010). Zonal cooperative inversion of crosshole p-wave, s-wave, and georadar traveltime data sets. \emph{Journal of Applied Geophysics}, 72(4), 254--262.

Marescot, L., Loke, M., Chapellier, D., Delaloye, R., Lambiel, C., \& Reynard, E. (2003). Assessing reliability of 2d resistivity imaging in mountain permafrost studies using the depth of investigation index method. \emph{Near Surface Geophysics}, 1(2), 57--67.

Marescot, L., Monnet, R., \& Chapellier, D. (2008). Resistivity and induced polarization surveys for slope instability studies in the swiss alps. \emph{Engineering Geology}, 98(1--2), 18--28.

Maurer, H., \& Hauck, C. (2007). Geophysical imaging of alpine rock glaciers. \emph{Journal of Glaciology}, 53(180), 110--120.

Meju, M. A., Gallardo, L. A., \& Mohamed, A. K. (2003). Evidence for correlation of electrical resistivity and seismic velocity in heterogeneous near-surface materials. \emph{Geophysical Research Letters}, 30(7), 1--4.

Mewes, B., Hilbich, C., Delaloye, R., \& Hauck, C. (2017). Resolution capacity of geophysical monitoring regarding permafrost degradation induced by hydrological processes. \emph{The Cryosphere}, 11(6), 2957--2974.

Mollaret, C., Wagner, F. M., Hilbich, C., Scapozza, C., \& Hauck, C. (2020). Petrophysical joint inversion applied to alpine permafrost field sites to image subsurface ice, water, air, and rock contents. \emph{Frontiers in Earth Science}, 8, 85. \url{https://doi.org/10.3389/feart.2020.00085}.

Monteiller, V. (2005). Tomographie à l'aide de décalages temporels d'ondes sismiques P: développements méthodologiques et applications. Université de Savoie. Ph.D. thesis.

Oggeri, C., \& Oreste, P. (2015). Underground quarrying for marble: Stability assessment through modelling and monitoring. \emph{International Journal of Mining Science (IJMS)}, 1(1), 35--42.

Oldenborger, G. A., Routh, P. S., \& Knoll, M. D. (2007). Model reliability for 3d electrical resistivity tomography: Application of the volume of investigation index to a time-lapse monitoring experiment. \emph{Geophysics}, 72(4), F167--F175.

Oldenburg, D. W., \& Li, Y. (1999). Estimating depth of investigation in dc resistivity and ip surveys. \emph{Geophysics}, 64(2), 403--416.

Parise, M., \& Lollino, P. (2011). A preliminary analysis of failure mechanisms in karst and man-made underground caves in southern Italy. \emph{Geomorphology}, 134(1), 132--143 (Geomorphology and Natural Hazards in Karst Areas).

Parsekian, A. D., Claes, N., Singha, K., Minsley, B. J., Carr, B., Voytek, E., et al. (2017). Comparing measurement response and inverted results of electrical resistivity tomography instruments. \emph{Journal of Environmental and Engineering Geophysics}, 22(3), 249--266.

Pedregosa, F., Varoquaux, G., Gramfort, A., Michel, V., Thirion, B., Grisel, O., et al. (2011). Scikit-learn: Machine learning in Python. \emph{Journal of Machine Learning Research}, 12, 2825--2830.

Pellet, C., Hilbich, C., Marmy, A., \& Hauck, C. (2016). Soil moisture data for the validation of permafrost models using direct and indirect measurement approaches at three alpine sites. \emph{Frontiers in Earth Science}, 3, 91.

Revil, A. (2013). Effective conductivity and permittivity of unsaturated porous materials in the frequency range 1 mHz--1 GHz. \emph{Water Resources Research}, 49(1), 306--327.

Revil, A., Cathles, L., III., Losh, S., \& Nunn, J. (1998). Electrical conductivity in shaly sands with geophysical applications. \emph{Journal of Geophysical Research: Solid Earth}, 103(B10), 23925--23936.

Revil, A., \& Linde, N. (2006). Chemico-electromechanical coupling in microporous media. \emph{Journal of Colloid and Interface Science}, 302(2), 682--694.

Revil, A., Schwaeger, H., Cathles, L., III., \& Manhardt, P. (1999). Streaming potential in porous media: 2. Theory and application to geothermal systems. \emph{Journal of Geophysical Research: Solid Earth}, 104(B9), 20033--20048.

Romero-Ruiz, A., Linde, N., Keller, T., \& Or, D. (2018). A review of geophysical methods for soil structure characterization. \emph{Reviews of Geophysics}, 56(4), 672--697.

Ronczka, M., Hellman, K., Günther, T., Wisén, R., \& Dahlin, T. (2017). Electric resistivity and seismic refraction tomography: a challenging joint underwater survey at äspö hard rock laboratory. \emph{Solid Earth}, 8(3), 671.

Rønning, J. S., Ganerød, G. V., Dalsegg, E., \& Reiser, F. (2014). Resistivity mapping as a tool for identification and characterisation of weakness zones in crystalline bedrock: Definition and testing of an interpretational model. \emph{Bulletin of Engineering Geology and the Environment}, 73(4), 1225--1244.

Rücker, C., Günther, T., \& Spitzer, K. (2006). Three-dimensional modelling and inversion of dc resistivity data incorporating topography-i modelling. \emph{Geophysical Journal International}, 166(2), 495--505.

Rücker, C., Günther, T., \& Wagner, F. M. (2017). pygimli: An open-source library for modelling and inversion in geophysics. \emph{Computers \& Geosciences}, 109, 106--123.

Srivastava, S., Pal, S. K., \& Kumar, R. (2020). A time-lapse study using self-potential and electrical resistivity tomography methods for mapping of old mine working across railway-tracks in a part of Raniganj coalfield, India. \emph{Environmental Earth Sciences}, 79(13), 1--19.

Tarantola, A. (2005). \emph{Inverse problem theory and methods for model parameter estimation}. SIAM.

Timur, A. (1968). Velocity of compressional waves in porous media at permafrost temperatures. \emph{Geophysics}, 33(4), 584--595.

Tritsch, J., Didier, C., \& Ghoreychi, M. (2007). Mise en sécurité des cavités souterraines d'origine anthropique: Surveillance-traitement. Guide technique. Rapport INERIS-DRS-07-86042-02484A, 15(02), 2007.

Tronicke, J., Holliger, K., Barrash, W., \& Knoll, M. D. (2004). Multivariate analysis of cross-hole georadar velocity and attenuation tomograms for aquifer zonation. \emph{Water Resources Research}, 40, W01519. \url{https://doi.org/10.1029/2003WR002031}.

Tsallis, C., Curado, E. M., de Souza, M., Elias, V. L., Bettini, C., Scuta, M. S., \& Beer, R. (1992). Generalized Archie law-application to petroleum reservoirs. \emph{Physica A: Statistical Mechanics and its Applications}, 191(1--4), 277--283.

Tso, C.-H. M., Kuras, O., Wilkinson, P. B., Uhlemann, S., Chambers, J. E., Meldrum, P. I., et al. (2017). Improved characterisation and modelling of measurement errors in electrical resistivity tomography (ERT) surveys. \emph{Journal of Applied Geophysics}, 146, 103--119.

Wagner, F., Mollaret, C., Günther, T., Kemna, A., \& Hauck, C. (2019). Quantitative imaging of water, ice and air in permafrost systems through petrophysical joint inversion of seismic refraction and electrical resistivity data. \emph{Geophysical Journal International}, 219(3), 1866--1875.

Waxman, M. H., Smits, L., et al. (1968). Electrical conductivities in oil-bearing shaly sands. \emph{Society of Petroleum Engineers Journal}, 8(02), 107--122.

Worthington, P. F. (2005). An electrical analog facility for hydrocarbon reservoirs. In: \emph{SPE Annual Technical Conference and Exhibition}. OnePetro.

Wyllie, M. R. J., Gregory, A. R., \& Gardner, L. W. (1956). Elastic wave velocities in heterogeneous and porous media. \emph{Geophysics}, 21(1), 41--70.

Xu, N., Li, T., Dai, F., Li, B., Zhu, Y., \& Yang, D. (2015). Microseismic monitoring and stability evaluation for the large scale underground caverns at the Houziyan hydropower station in southwest China. \emph{Engineering Geology}, 188, 48--67.

\endgroup

\end{document}